\documentclass[preprint]{elsarticle}

\usepackage{moreverb}
\usepackage[svgnames=true]{xcolor}
\usepackage{graphicx}
\usepackage[misc]{ifsym}

\usepackage{tabularx}
\usepackage{adjustbox}
\usepackage{rotating}

\usepackage{algorithmicx}
\usepackage[noend]{algpseudocode}
\usepackage{algorithm}

\usepackage{url}
\usepackage{xspace}

\usepackage{amsmath}
\usepackage{amsfonts}
\usepackage{amsfonts}
\usepackage{mathrsfs}

\usepackage{booktabs}
\usepackage{float}
\usepackage[]{microtype}

\usepackage{proof}
\usepackage{paralist}

\usepackage[binary-units=true]{siunitx}

\usepackage{ifthen}

\usepackage{times}
\usepackage{tablefootnote}

\usepackage{paralist}

\usepackage{fontawesome}
\usepackage{multirow}

\usepackage{subcaption}
\usepackage{wrapfig}
\usepackage{framed}
\usepackage[pscoord]{eso-pic} % The zero point of the coordinate system is the lower left corner of the page (the default).
\usepackage{svg}

\usepackage{listings}
\usepackage{array}
\usepackage{mdwmath}
\usepackage{mdwtab}

\usepackage{stfloats}

\usepackage{enumerate}

\newcommand{\bs}{client/server code splitting\xspace}
\newcommand{\cs}{\bs}

\usepackage[normalem]{ulem}
\usepackage[acronym,nomain]{glossaries}
\glsdisablehyper
\newcommand{\changed}[1]{{#1}}

\newcommand{\changeda}[1]{{#1}}
\newcommand{\ie}{i.e.\@\xspace} % Id est.
\newcommand{\eg}{e.g.\@\xspace} % Exempli gratia.
\newcommand{\wrt}{w.r.t.\@\xspace} % With respect to.
\newcommand{\etal}{\textit{~et~al.\@}\xspace}

\newacronym{nist}{NIST}{National Institute of Standards and Technology}
\newacronym{gdpr}{GDPR}{General Data Protection Regulation}

\newacronym{asp}{ASPIRE}{Advanced Software Protection: Integration, Research and Exploitation}
\newacronym{accl}{ACCL}{Client-side Communication Logic}
\newacronym{ascl}{ASCL}{Server-side Communication Logic}
\newacronym{actc}{ACTC}{ASPIRE Compiler Tool Chain}
\newacronym{esp}{ESP}{Expert system for Software Protection}
\newacronym{it}{IT}{Information Technology}

\newacronym{diablo}{DIABLO}{Diablo Is A Better Link-time Optimizer}
\newacronym{cff}{CFF}{Control Flow Flattening}\newcommand{\cff}{\gls{cff}\xspace}
\newacronym{op}{OP}{Opaque Predicate}
\newacronym{ops}{OPs}{Opaque Predicates}\newcommand{\ops}{\gls{ops}\xspace}
\newacronym{api}{API}{Application Programming Interface}
\newacronym{ip}{IP}{Intellectual Property}
\newacronym{drm}{DRM}{Digital Rights Management}
\newacronym{cve}{CVE}{Common Vulnerabilities and Exposures}
\newacronym{cwe}{CWE}{Common Weakness Enumeration}
\newacronym{mate}{MATE}{Man-At-The-End}\newcommand{\mate}{\gls{mate}\xspace}
\newacronym{sp}{SP}{Software Protection}

\newacronym{txt}{TXT}{Trusted eXecution Technology}
\newacronym{tpm}{TPM}{Trusted Platform Module}
\newacronym{tcg}{TCG}{Trusted Computing Group}
\newacronym{sgx}{SGX}{Software Guard Extensions}

\newacronym{dec}{DEC}{Digital Equipment Corporation}
\newacronym{vax}{VAX}{Virtual Address eXtension}

\newacronym{dendral}{DENDRAL}{DENDritic ALgorithm}
\newacronym{xcon}{XCON}{eXpert CONfigurer}
\newacronym{ides}{IDES}{Intrusion Detection Expert System}
\newacronym{nides}{NIDES}{Next-generation Intrusion Detection Expert System}
\newacronym{nidx}{NIDX}{Network Intrusion Detection eXpert system}
\newacronym{nadir}{NADIR}{Network Anomaly Detection and Intrusion Reporter}
\newacronym{audes}{AudES}{Expert System for security Auditing}

\newacronym{cfg}{CFG}{Control Flow Graph}\newcommand{\cfg}{\gls{cfg}\xspace}
\newacronym{os}{OS}{Operating System}
\newacronym{dll}{DLL}{Dynamic-Link Library}
\newacronym{gnu}{GNU}{GNU is Not Unix}
\newacronym{gcc}{GCC}{\gls{gnu} C Compiler}
\newacronym{gdb}{GDB}{\gls{gnu} DeBugger}
\newacronym{foss}{FOSS}{Free and Open Source Software}
\newacronym{cdt}{CDT}{C/C++ Development Tooling}
\newacronym{ast}{AST}{Abstract Syntax Tree}
\newacronym{gui}{GUI}{Graphical User Interface}
\newacronym{xml}{XML}{eXtensible Mark-up Language}

\newacronym{sloc}{SLOC}{Source Lines Of Code}\newcommand{\sloc}{\gls{sloc}\xspace}
\newacronym{hl}{HL}{Halstead's Length}
\newacronym{cc}{CC}{Cyclomatic Complexity}\newcommand{\CC}{\gls{cc}\xspace}

\newacronym{cpu}{CPU}{Central Processing Unit}
\newacronym{fpu}{FPU}{Floating Point Unit}
\newacronym{alu}{ALU}{Arithmetic and Logic Unit}
\newacronym{gpu}{GPU}{Graphics Processing Unit}
\newacronym{gpgpu}{GPGPU}{General-Purpose computing on Graphics Processing Units}

\newacronym{lisp}{LISP}{LISt Processor}
\newacronym{cil}{CIL}{C Intermediate Language}

\newacronym{smb}{SMB}{Server Message Block}

\newacronym{mitm}{MITM}{Man-In-The-Middle}
\newacronym{ddos}{DDoS}{Distributed Denial of Service}

\newacronym{rat}{RAT}{Remote Access Trojan}

\newacronym{ids}{IDS}{Intrusion Detection System}
\newacronym{cysemol}{CySeMoL}{Cyber Security Modeling Language}
\newacronym{camel}{CAMEL}{Cloud Application Modelling \& Execution Language}
\newacronym{ml}{ML}{Machine Learning}
\newacronym{ai}{AI}{Artificial Intelligence}\newcommand{\ai}{\gls{ai}\xspace}
\newacronym{som}{SOM}{Self-Organizing Maps}
\newacronym{ann}{ANN}{Artificial Neural Network}
\newacronym{llm}{LLM}{Large Language Model}
\newacronym{llms}{LLMs}{Large Language Models}\newcommand{\llms}{\gls{llms}\xspace}

\newacronym{milp}{MILP}{Mixed Integer-Linear Programming}

\newacronym{owl}{OWL}{Web Ontology Language}
\newacronym{owltwo}{OWL2}{\acrlong{owl} 2}

\newacronym{roc}{ROC}{Receiver Operating Characteristic}\newcommand{\roc}{\gls{roc}\xspace}
\newacronym{auc}{AUC}{Area Under the Curve}\newcommand{\auc}{\gls{auc}\xspace}

\usepackage{tikz}
\usetikzlibrary{
	decorations,
	arrows,
	shapes,positioning,
	patterns,
	fit,
	automata,
	decorations.markings,
	plotmarks,
	decorations.pathmorphing,
	positioning,
	backgrounds,
	fit,
	calc,
	arrows.meta}

\usepackage{pgfplots}
\pgfplotsset{ cycle list={%
		{smooth,black,mark=triangle*},
		{smooth,black,mark=triangle},
		{smooth,black,mark=*},
		{smooth,black,mark=o},
		{smooth,black,mark=square*},
		{smooth,black,mark=square},
		{smooth,black,mark=diamond*},
		{smooth,black},
		{smooth,black},
		{smooth,black},
		{smooth,black},
		{smooth,black},
		{smooth,black},
		{smooth,black},
		{smooth,black},
		{smooth,black},
		{smooth,black},
		{smooth,black},
	},
	compat=1.3,
	tick label style={ font=\scriptsize,/pgf/number format/set thousands separator={}}}
\usepgfplotslibrary{fillbetween}

\usepackage{hyperref}

\newcolumntype{L}{>{\raggedright\arraybackslash}X}

\newcommand{\VersionCFF}{$C\!F\!F_s$\xspace}
\newcommand{\VersionCFFOp}{$C\!F\!F_{op}$\xspace}
\newcommand{\VersionDoubleCFF}{$C\!F\!F_s+C\!F\!F_{op}$\xspace}
\newcommand{\VersionCFFMATH}{C\!F\!F_s}
\newcommand{\VersionCFFOpMATH}{C\!F\!F_{op}}
\newcommand{\VersionDoubleCFFMATH}{C\!F\!F_s+C\!F\!F_{op}}

\newcommand{\AppA}{$\mathcal{A}_a$\xspace}
\newcommand{\AppN}{$\mathcal{A}_n$\xspace}
\newcommand{\AppT}{$\mathcal{A}_t$\xspace}

\newbox\@BILD%
\newbox\@TEXT%
\newdimen\d@breite%
\newdimen\d@hoehe%
\newdimen\d@xoff%
\newdimen\d@yoff%
\newdimen\d@shad%
\newdimen\d@dash%
\newdimen\d@boxl%
\newdimen\d@pichskip%
\newdimen\d@tmp
\newdimen\d@tmpa
\newdimen\d@bskip
\newdimen\hsiz@%
\newdimen\p@getot@l%
\newcount\c@breite
\newcount\c@hoehe
\newcount\c@xoff
\newcount\c@yoff
\newcount\c@pos
\newcount\c@shad
\newcount\c@dash
\newcount\c@boxl
\newcount\c@zeilen%
\newcount\@changemode%
\newcount\c@piccaption%
\newcount\c@piccaptionpos%
\newcount\c@picpos
\newcount\c@whole%
\newcount\c@half%
\newcount\c@tmp
\newcount\c@tmpa
\newcount\c@tmpb
\newcount\c@tmpc
\newcount\c@tmpd
\newskip\d@leftskip
\newif\if@list \@listfalse%
\newif\if@offset%

\def\@captype{figure}%
\let\old@par=\par%

\def\pichskip#1{\d@pichskip #1\relax}

\def\shadowthickness#1{\d@shad #1\relax}

\def\dashlength#1{\d@dash #1\relax}

\def\boxlength#1{\d@boxl #1\relax}

\def\picchangemode{\@changemode=1}%
\def\nopicchangemode{\@changemode=0}%

\def\piccaptionoutside{\c@piccaptionpos=1}%
\def\piccaptioninside{\c@piccaptionpos=2}%
\def\piccaptionside{\c@piccaptionpos=3}%
\def\piccaptiontopside{\c@piccaptionpos=4}%

\def\piccaption{\@ifnextchar [{\@piccaption}{\@piccaption[]}}
\def\@piccaption[#1]#2{\c@piccaption=1\def\sh@rtf@rm{#1}\def\capti@nt@xt{#2}}
\def\make@piccaption{%
 \hsiz@\d@breite%
 \ifnum\c@piccaptionpos=2%
   \advance\hsiz@ -2\fboxsep%
 \fi%
 \ifnum\c@piccaptionpos>2%
   \hsiz@\hsize\advance\hsiz@-\d@breite\advance\hsiz@-\d@pichskip%
 \fi%
 \setbox\@TEXT=\vbox{\hsize\hsiz@\caption[\sh@rtf@rm]{\capti@nt@xt}}%
}

\def\newcaption{\refstepcounter\@captype\@dblarg{\@newcaption\@captype}}
\long\def\@newcaption#1[#2]#3{%
  \old@par%
  \addcontentsline{\csname ext@#1\endcsname }{#1}%
    {\protect\numberline{\csname the#1\endcsname}{\ignorespaces #2}}
  \begingroup\@parboxrestore\normalsize%
    \@newmakecaption{\csname fnum@#1\endcsname}{\ignorespaces #3}\old@par%
  \endgroup%
}
\long\def\@newmakecaption#1#2{%
  \vskip 10pt%
  \setbox\@tempboxa \hbox {#1: #2}%
  \ifdim \wd\@tempboxa >\hsize%
    \setbox0=\hbox{#1: }\dimen0=\hsize\advance\dimen0 by-\wd0
    \setbox1=\vtop{\hsize=\dimen0 #2}
    \hbox{\box0 \box1}
    \par
  \else \hbox to\hsize {\hfil \box \@tempboxa \hfil}
  \fi
}

\def\parpic{%
  \@ifnextchar ({\iparpic}{\iparpic(0pt,0pt)}
}
\def\iparpic(#1,#2){%
  \@ifnextchar ({\@offsettrue\iiparpic(#1,#2)}%
                {\@offsetfalse\iiparpic(#1,#2)(0pt,0pt)}
}
\def\iiparpic(#1,#2)(#3,#4){%
  \@ifnextchar [{\iiiparpic(#1,#2)(#3,#4)}{\iiiparpic(#1,#2)(#3,#4)[l]}
}
\def\iiiparpic(#1,#2)(#3,#4)[#5]{%
  \@ifnextchar [{\ivparpic(#1,#2)(#3,#4)[#5]}{\ivparpic(#1,#2)(#3,#4)[#5][]}
}
\def\ivparpic(#1,#2)(#3,#4)[#5][#6]#7{%
 \let\par=\old@par\par%
 \hangindent0pt\hangafter1%
 \setbox\@BILD=\hbox{#7}%
 \d@breite=#1\d@breite=\the\d@breite%
 \ifdim\d@breite=0pt\d@breite=\wd\@BILD\fi%
 \c@breite=\d@breite\divide\c@breite by65536%
 \multiply\c@piccaption\c@piccaptionpos%
 \d@hoehe=#2\d@hoehe=\the\d@hoehe%
 \ifdim\d@hoehe=0pt\d@hoehe=\ht\@BILD\advance\d@hoehe by\dp\@BILD\fi%
 \c@hoehe=\d@hoehe\divide\c@hoehe by65536%
 \d@xoff=#3\c@xoff=\d@xoff\divide\c@xoff by65536%
 \d@yoff=\d@hoehe%
 \advance\d@yoff by-#4\c@yoff=\d@yoff\divide\c@yoff by65536%
 \c@pos=1\unitlength1pt%
 \if@offset%
   \setbox\@BILD=\hbox{%
     \begin{picture}(\c@breite,\c@hoehe)%
       \put(0,0){\makebox(\c@breite,\c@hoehe){}}%
       \put(\c@xoff,\c@yoff){\box\@BILD}%
     \end{picture}%
   }%
 \else%
   \setbox\@BILD=\hbox{%
     \begin{picture}(\c@breite,\c@hoehe)%
       \put(0,0){\makebox(\c@breite,\c@hoehe)[#6]{\box\@BILD}}%
     \end{picture}%
   }%
 \fi%
 \ifnum\c@piccaption=2%
   \make@piccaption%
   \advance\d@hoehe\ht\@TEXT\advance\d@hoehe\dp\@TEXT%
   \c@hoehe=\d@hoehe\divide\c@hoehe by65536%
   \setbox\@BILD=\vbox{\box\@BILD\vspace{-5pt}%
                       \hbox{\hspace{\fboxsep}\box\@TEXT}%
                       \vspace{4pt}}%
 \fi%
 \@tfor\@tempa := #5\do{%
   \if\@tempa f\setbox\@BILD=\hbox{\Rahmen(\c@breite,\c@hoehe){\box\@BILD}}\fi%
   \if\@tempa s\setbox\@BILD=\hbox{\Schatten(\c@breite,\c@hoehe){\box\@BILD}}\fi%
   \if\@tempa o\setbox\@BILD=\hbox{\Oval(\c@breite,\c@hoehe){\box\@BILD}}\fi%
   \if\@tempa d\setbox\@BILD=\hbox{\Strich(\c@breite,\c@hoehe){\box\@BILD}}\fi%
   \if\@tempa x\setbox\@BILD=\hbox{\Kasten(\c@breite,\c@hoehe){\box\@BILD}}\fi%
   \if\@tempa l\c@pos=1\fi%
   \if\@tempa r\c@pos=2\fi%
 }%
 \ifnum\c@piccaption=1%
   \make@piccaption%
   \advance\d@hoehe\ht\@TEXT\advance\d@hoehe\dp\@TEXT%
   \c@hoehe=\d@hoehe\divide\c@hoehe by65536%
   \setbox\@BILD=\vbox{\box\@BILD\vspace{-5pt}\hbox{\box\@TEXT}\vspace{4pt}}%
 \fi%
 \ifodd\count0\c@picpos=0\else\c@picpos=\@changemode\fi%
 \pagetotal=\the\pagetotal%
 \d@tmp=\pagegoal\advance\d@tmp by-\pagetotal\advance\d@tmp by-\baselineskip%
 \ifdim\d@hoehe>\d@tmp%
   \vskip 0pt plus\d@hoehe\relax\pagebreak[3]\vskip 0pt plus-\d@hoehe\relax%
   \ifnum\c@picpos=1\c@picpos=0\else\c@picpos=\@changemode\fi%
 \fi%
 \ifnum\c@picpos=1\ifnum\c@pos=1\c@pos=2\else\c@pos=1\fi\fi%
 \ifnum\@listdepth>0
   \@listtrue\parshape 0%
   \advance\hsize -\rightmargin%
   \d@leftskip \leftskip%
   \leftskip \@totalleftmargin%
   \if@inlabel\rule{\linewidth}{0pt}\vskip-\baselineskip\relax\fi%
 \else\@listfalse\medskip%
 \fi%
 \if@list\d@tmpa=\linewidth\else\d@tmpa=\hsize\fi%
 \ifnum\c@piccaption=3%
   \make@piccaption%
   \d@tmp\ht\@TEXT\advance\d@tmp\dp\@TEXT%
   \ifdim\d@hoehe>\d@tmp%
     \setbox\@TEXT=\vbox to\d@hoehe{\vfill\box\@TEXT\vspace{.2\baselineskip}\vfill}%
   \else%
     \setbox\@BILD=\vbox to\d@tmp{\vfill\box\@BILD\vfill}%
     \d@hoehe\d@tmp%
   \fi%
 \fi%
 \ifnum\c@piccaption=4%
   \make@piccaption%
   \d@tmp\ht\@TEXT\advance\d@tmp\dp\@TEXT%
   \setbox\@TEXT=\vbox to\d@hoehe{\vspace{-10pt}\box\@TEXT\vfil}%
   \advance\d@hoehe-\d@tmp%
 \fi%
 \ifnum\c@pos=1\d@tmpa=0pt%
   \ifnum\c@piccaption>2%
      \setbox\@BILD=\hbox{\box\@BILD\hspace{\d@pichskip}\hbox{\box\@TEXT}}%
   \fi%
 \else\advance\d@tmpa by-\wd\@BILD\d@breite=-\d@breite%
   \ifnum\c@piccaption>2%
      \d@tmpa=0pt%
      \setbox\@BILD=\hbox{\hbox{\box\@TEXT}\hspace{\d@pichskip}\box\@BILD}%
   \fi%
 \fi%
 \p@getot@l\the\pagetotal%
 \d@bskip\d@hoehe\advance\d@bskip by\parskip\advance\d@bskip by.3\baselineskip%
 {\noindent\hspace*{\d@tmpa}\relax%
  \box\@BILD\nopagebreak\vskip-\d@bskip\relax\nopagebreak}%
 \d@tmp=-\d@hoehe\divide\d@tmp by\baselineskip%
 \c@zeilen=\d@tmp\advance\c@zeilen by-1%
 \ifdim\d@breite<0pt\advance\d@breite by-\d@pichskip%
 \else\advance\d@breite by\d@pichskip%
 \fi%
 \hangindent=\d@breite%
 \hangafter=\c@zeilen%
 \let\par=\x@par%
 \ifnum\c@piccaption=3%
    \hangindent0pt\hangafter1\let\par=\old@par%
    \vskip\d@hoehe\vskip.2\baselineskip%
 \fi%
 \c@piccaption=0%
}

\newdimen\ptoti
\newdimen\ptotii
\def\x@par{%
 \ptoti\pagetotal%
 \old@par%
 \ptotii\pagetotal%
 \ifdim\ptoti=\ptotii%
   \d@tmp\d@hoehe%
 \else%
   \d@tmp\baselineskip%
   \multiply\d@tmp by\prevgraf%
   \advance\d@tmp by\parskip%
   \global\advance\d@hoehe by-\d@tmp\d@tmp=\d@hoehe%
 \fi%
 \ifdim\d@hoehe>0pt%
   \divide\d@tmp by\baselineskip\c@zeilen=-\d@tmp\advance\c@zeilen by-1%
   \c@zeilen=\the\c@zeilen%
 \else\c@zeilen=0
 \fi
 \ifnum\c@zeilen<0\hangafter=\c@zeilen\hangindent=\d@breite%
 \else\let\par=\old@par%
   \hangindent 0pt%
   \leftskip \d@leftskip%
   \if@list\parshape \@ne \@totalleftmargin \linewidth%
     \advance\hsize \rightmargin%
   \fi%
 \fi%
}

\def\picskip#1{%
 \let\par=\old@par%
 \par%
 \pagetotal\the\pagetotal%
 \c@tmp=#1\relax%
 \ifnum\c@tmp=0%
   \d@tmp\baselineskip\multiply\d@tmp by\prevgraf\advance\d@tmp\parskip%
   \ifdim\p@getot@l<\pagetotal
     \advance\d@hoehe by-\d@tmp\advance\d@hoehe by1ex%
     \ifdim\d@hoehe>0pt\vspace*{\d@hoehe}\fi%
   \fi%
   \ifdim\p@getot@l=\pagetotal%
     \advance\d@hoehe by-\d@tmp\advance\d@hoehe by1ex%
     \ifdim\d@hoehe>0pt\vspace*{\d@hoehe}\fi%
   \fi%
 \else\hangafter=-\c@tmp\hangindent=\d@breite%
 \fi%
 \leftskip \d@leftskip%
 \if@list\parshape \@ne \@totalleftmargin \linewidth%
   \advance\hsize \rightmargin%
 \fi%
}

\def\hpic{%
  \@ifnextchar ({\ihpic}{\ihpic(0pt,0pt)}
}
\def\ihpic(#1,#2){%
  \@ifnextchar ({\@offsettrue\iihpic(#1,#2)}%
                {\@offsetfalse\iihpic(#1,#2)(0pt,0pt)}
}
\def\iihpic(#1,#2)(#3,#4){%
  \@ifnextchar [{\iiihpic(#1,#2)(#3,#4)}{\iiihpic(#1,#2)(#3,#4)[l]}
}
\def\iiihpic(#1,#2)(#3,#4)[#5]{%
  \@ifnextchar [{\ivhpic(#1,#2)(#3,#4)[#5]}{\ivhpic(#1,#2)(#3,#4)[#5][]}
}
\def\ivhpic(#1,#2)(#3,#4)[#5][#6]#7{%
  \setbox\@BILD=\hbox{#7}%
  \d@breite=#1\d@breite=\the\d@breite%
  \ifdim\d@breite=0pt\d@breite=\wd\@BILD\fi%
  \c@breite=\d@breite\divide\c@breite by65536%
  \d@hoehe=#2\d@hoehe=\the\d@hoehe%
  \ifdim\d@hoehe=0pt\d@hoehe=\ht\@BILD\advance\d@hoehe by\dp\@BILD\fi%
  \c@hoehe=\d@hoehe\divide\c@hoehe by65536%
  \d@xoff=#3\c@xoff=\d@xoff\divide\c@xoff by65536%
  \d@yoff=\d@hoehe%
  \advance\d@yoff by-#4\c@yoff=\d@yoff\divide\c@yoff by65536%
  \c@pos=0\d@tmpa=\parindent\parindent=0pt\unitlength1pt%
  \if@offset
    \setbox\@BILD=\hbox{%
      \begin{picture}(\c@breite,\c@hoehe)%
        \put(0,0){\makebox(\c@breite,\c@hoehe){}}%
        \put(\c@xoff,\c@yoff){\box\@BILD}%
      \end{picture}%
    }%
  \else%
    \setbox\@BILD=\hbox{%
      \begin{picture}(\c@breite,\c@hoehe)%
        \put(0,0){\makebox(\c@breite,\c@hoehe)[#6]{\box\@BILD}}%
      \end{picture}%
    }%
  \fi%
  \@tfor\@tempa := #5\do{%
    \if\@tempa f\setbox\@BILD=\hbox{\Rahmen(\c@breite,\c@hoehe){\box\@BILD}}\fi%
    \if\@tempa s\setbox\@BILD=\hbox{\Schatten(\c@breite,\c@hoehe){\box\@BILD}}\fi%
    \if\@tempa o\setbox\@BILD=\hbox{\Oval(\c@breite,\c@hoehe){\box\@BILD}}\fi%
    \if\@tempa d\setbox\@BILD=\hbox{\Strich(\c@breite,\c@hoehe){\box\@BILD}}\fi%
    \if\@tempa x\setbox\@BILD=\hbox{\Kasten(\c@breite,\c@hoehe){\box\@BILD}}\fi%
    \if\@tempa t\c@pos=1\fi%
    \if\@tempa b\c@pos=2\fi%
  }%
 \ifnum\c@pos=0\parbox{\d@breite}{\makebox[0cm]{}\\\box\@BILD\smallskip}\fi%
 \ifnum\c@pos=1\parbox[t]{\d@breite}{\makebox[0cm]{}\\\box\@BILD\smallskip}\fi%
 \ifnum\c@pos=2\parbox[b]{\d@breite}{\makebox[0cm]{}\\\box\@BILD\smallskip}\fi%
 \parindent=\d@tmpa%
}

\def\Rahmen(#1,#2)#3{%
  \c@whole=\@wholewidth\divide\c@whole by65536%
  \c@half=\@halfwidth\divide\c@half by65536%
  \c@tmpa=#1\advance\c@tmpa by\c@whole\advance\c@tmpa by\c@whole%
  \c@tmpb=#2\advance\c@tmpb by\c@whole\advance\c@tmpb by\c@whole%
  \begin{picture}(\c@tmpa,\c@tmpb)%
    \put(\c@whole,\c@half){\framebox(#1,#2){#3}}%
  \end{picture}%
  \global\advance\d@breite by2\@wholewidth%
  \global\advance\d@hoehe by2\@wholewidth%
}

\def\Schatten(#1,#2)#3{%
  \c@whole=\@wholewidth\divide\c@whole by65536%
  \c@half=\@halfwidth\divide\c@half by65536%
  \c@shad=\d@shad\divide\c@shad by65536%
  \c@tmp=\c@whole\advance\c@tmp by\c@whole\c@tmpd=\c@tmp%
  \advance\c@tmp by\c@shad%
  \advance\c@tmpd by#1%
  \advance\c@half by\c@shad%
  \c@tmpa=#1\advance\c@tmpa by\c@tmp%
  \c@tmpb=#2\advance\c@tmpb by\c@tmp%
  \begin{picture}(\c@tmpa,\c@tmpb)%
    \put(\c@whole,\c@half){\framebox(#1,#2){#3}}%
    \put(\c@shad,0){\rule{\c@tmpd pt}{\c@shad pt}}%
    \put(\c@tmpd,0){\rule{\c@shad pt}{#2 pt}}%
  \end{picture}%
  \global\advance\d@breite by2\@wholewidth\global\advance\d@breite by\d@shad%
  \global\advance\d@hoehe by2\@wholewidth\global\advance\d@hoehe by\d@shad%
}

\def\Oval(#1,#2)#3{%
  \@wholewidth=0.4pt%
  \c@tmpa=\the#1\divide\c@tmpa by2%
  \c@tmpb=\the#2\divide\c@tmpb by2%
  \begin{picture}(#1,#2)%
    \put(\c@tmpa,\c@tmpb){\oval(#1,#2)}%
    \put(0.4,0.4){#3}%
  \end{picture}%
  \global\advance\d@breite by1pt\global\advance\d@hoehe by1pt%
}

\def\Strich(#1,#2)#3{%
  \c@whole=\@wholewidth\divide\c@whole by65536%
  \c@half=\@halfwidth\divide\c@half by65536%
  \c@dash=\d@dash\divide\c@dash by65536%
  \c@tmp=\c@whole\advance\c@tmp by\c@whole%
  \c@tmpa=#1\advance\c@tmpa by\c@tmp%
  \c@tmpb=#2\advance\c@tmpb by\c@tmp%
  \c@tmpc=#1\advance\c@tmpc by\c@whole%
  \c@tmpd=#2\advance\c@tmpd by\c@whole%
  \begin{picture}(\c@tmpa,\c@tmpb)%
    \put(\c@half,\c@half){\dashbox{\c@dash}(\c@tmpc,\c@tmpd){#3}}%
  \end{picture}%
  \global\advance\d@breite by2\@wholewidth%
  \global\advance\d@hoehe by2\@wholewidth%
}

\def\Kasten(#1,#2)#3{%
  \@wholewidth=0.4pt%
  \c@boxl=\d@boxl\divide\c@boxl by65536\c@boxl=\the\c@boxl%
  \c@tmpa=#1\advance\c@tmpa by\c@boxl%
  \c@tmpb=#2\advance\c@tmpb by\c@boxl%
  \c@tmp=#2%
  \begin{picture}(\c@tmpa,\c@tmpb)%
    \put(0,\c@boxl){\framebox(#1,#2){#3}}%
    \put(\c@boxl,0){\line(-1,1){\c@boxl}}%
    \put(\c@boxl,0){\line(1,0){#1}\line(-1,1){\c@boxl}}%
    \put(\c@boxl,0){\put(#1,0){\line(0,1){\c@tmp}%
         \put(0,\c@tmp){\line(-1,1){\c@boxl}}}}%
  \end{picture}%
  \global\advance\d@breite by\d@boxl%
  \global\advance\d@hoehe by\d@boxl%
}

\newbox\env@box%
\newdimen\d@envdp
\newcount\c@hsize
\newcount\c@envdp
\newdimen\d@envb

\long\def\frameenv{\@ifnextchar [{\@frameenv}{\@frameenv[\textwidth]}}
\long\def\@frameenv[#1]{%
 \hsiz@=\textwidth  \textwidth=#1  \d@envb=#1
 \advance\textwidth by-2\@wholewidth
 \advance\textwidth by-2\fboxsep
 \hsize=\textwidth   \linewidth=\textwidth
 \setbox\env@box=\vbox\bgroup}%
\def\endframeenv{%
 \egroup%
 \hsize=\hsiz@  \textwidth=\hsiz@  \linewidth=\hsiz@
 \c@breite=\d@envb   \divide\c@breite by65536
 \advance\d@envb by-2\@wholewidth
 \c@hsize=\d@envb  \divide\c@hsize by65536%
 \d@envdp=\dp\env@box  \advance\d@envdp by\ht\env@box%
 \advance\d@envdp by2\fboxsep%
 \d@hoehe=\d@envdp   \advance\d@hoehe by2\@wholewidth
 \c@hoehe=\d@hoehe   \divide\c@hoehe by65536
 \c@envdp=\d@envdp   \divide\c@envdp by65536%
 \c@tmp=\@wholewidth \divide\c@tmp by65536
 \vskip\@wholewidth%
 \unitlength 1pt\noindent%
 \begin{picture}(\c@breite,\c@hoehe)(0,0)
   \put(\c@tmp,\c@tmp){\framebox(\c@hsize,\c@envdp){\box\env@box}}
 \end{picture}%
}

\long\def\shadowenv{\@ifnextchar [{\@shadowenv}{\@shadowenv[\textwidth]}}
\long\def\@shadowenv[#1]{%
 \hsiz@=\textwidth  \textwidth=#1  \d@envb=#1
 \advance\textwidth by-2\@wholewidth
 \advance\textwidth by-2\fboxsep
 \advance\textwidth by-\d@shad%
 \hsize=\textwidth   \linewidth=\textwidth
 \setbox\env@box=\vbox\bgroup}%
\def\endshadowenv{%
 \egroup
 \hsize=\hsiz@  \textwidth=\hsiz@  \linewidth=\hsiz@
 \d@tmpa=\d@envb
 \c@breite=\d@envb   \divide\c@breite by65536
 \advance\d@envb by-2\@wholewidth  \advance\d@envb by-\d@shad
 \c@hsize=\d@envb  \divide\c@hsize by65536%
 \d@envdp=\dp\env@box  \advance\d@envdp by\ht\env@box%
 \advance\d@envdp by2\fboxsep%
 \c@envdp=\d@envdp   \divide\c@envdp by65536%
 \d@hoehe=\d@envdp
 \advance\d@hoehe by2\@wholewidth  \advance\d@hoehe by\d@shad
 \c@hoehe=\d@hoehe    \divide\c@hoehe by65536
 \c@shad =\d@shad     \divide\c@shad  by65536
 \c@tmp=\@wholewidth  \divide\c@tmp by65536
 \advance\d@tmpa by-2\d@shad
 \c@xoff =\d@tmpa     \divide\c@xoff by65536
 \advance\c@xoff by\c@shad  \advance\c@xoff by-1
 \advance\d@envdp by\@wholewidth
 \vskip\@halfwidth
 \unitlength 1pt\noindent%
 \begin{picture}(\c@breite,\c@hoehe)(0,0)
    \put(\c@tmp,\c@shad){\framebox(\c@hsize,\c@envdp){\box\env@box}}
    \put(\c@shad,0){\rule{\d@tmpa}{\d@shad}}%
    \put(\c@xoff,0){\rule{\d@shad}{\d@envdp}}%
 \end{picture}%
 \vskip\@halfwidth
}

\long\def\dashenv{\@ifnextchar [{\@dashenv}{\@dashenv[\textwidth]}}
\long\def\@dashenv[#1]{%
 \hsiz@=\textwidth  \textwidth=#1  \d@envb=#1
 \advance\textwidth by-2\@wholewidth  \advance\textwidth by-2\fboxsep
 \hsize=\textwidth   \linewidth=\textwidth
 \setbox\env@box=\vbox\bgroup}%
\long\def\enddashenv{%
 \egroup
 \hsize=\hsiz@  \textwidth=\hsiz@  \linewidth=\hsiz@
 \c@breite=\d@envb   \divide\c@breite by65536
 \advance\d@envb by-\@wholewidth
 \c@hsize=\d@envb  \divide\c@hsize by65536%
 \d@envdp=\dp\env@box  \advance\d@envdp by\ht\env@box%
 \advance\d@envdp by2\fboxsep%
 \advance\d@envdp by\@wholewidth
 \d@hoehe=\d@envdp   \advance\d@hoehe by2\@wholewidth
 \c@hoehe=\d@hoehe   \divide\c@hoehe by65536
 \c@envdp=\d@envdp   \divide\c@envdp by65536%
 \c@dash=\d@dash     \divide\c@dash  by65536%
 \c@whole=\@wholewidth  \divide\c@whole by65536
 \c@half=\@halfwidth  \divide\c@half by 65536
 \noindent\unitlength 1pt
 \begin{picture}(\c@breite,\c@hoehe)(0,0)
   \put(\c@half,\c@whole){\dashbox{\c@dash}(\c@hsize,\c@envdp){\box\env@box}}
 \end{picture}%
}

\long\def\ovalenv{\@ifnextchar [{\@ovalenv}{\@ovalenv[\textwidth]}}%
\long\def\@ovalenv[#1]{%
 \hsiz@=\textwidth  \textwidth=#1  \d@envb=#1
 \advance\textwidth by-4\fboxsep
 \hsize=\textwidth   \linewidth=\textwidth
 \setbox\env@box=\vbox\bgroup}%
\long\def\endovalenv{%
 \egroup
 \hsize=\hsiz@  \textwidth=\hsiz@  \linewidth=\hsiz@
 \@wholewidth=0.4pt
 \c@breite=\d@envb   \divide\c@breite by65536
 \advance\d@envb by-2\@wholewidth
 \c@hsize=\d@envb  \divide\c@hsize by65536%
 \d@envdp=\dp\env@box  \advance\d@envdp by\ht\env@box%
 \advance\d@envdp by4\fboxsep%
 \c@envdp=\d@envdp   \divide\c@envdp by65536%
 \d@hoehe=\d@envdp   \advance\d@hoehe by2\@wholewidth
 \c@hoehe=\d@hoehe   \divide\c@hoehe by65536
 \c@tmpa=\c@hsize   \divide\c@tmpa by2%
 \c@tmpb=\c@envdp   \divide\c@tmpb by2%
 \d@tmpa=2\fboxsep   \advance\d@tmpa by\@wholewidth
 \c@xoff=\d@tmpa     \divide\c@xoff  by65536%
 \advance\d@tmpa by\dp\env@box
 \c@yoff=\d@tmpa     \divide\c@yoff  by65536%
 \unitlength 1pt\noindent
 \begin{picture}(\c@breite,\c@hoehe)(0,0)
   \put(\c@tmpa,\c@tmpb){\oval(\c@hsize,\c@envdp)}
   \put(\c@xoff,\c@yoff){\box\env@box}%
 \end{picture}%
}

\journal{Computers \& Security}

\begin{document}

\begin{frontmatter}
  
\title{Empirical Assessment of the \changed{Code Comprehension} Effort Needed to Attack Programs Protected with Obfuscation}

\author[1]{Leonardo~Regano\corref{cor1}}
\ead{leonardo.regano@unica.it}
\author[2]{Daniele~Canavese}
\ead{daniele.canavese@cnr.it}
\author[3]{Cataldo~Basile}
\ead{cataldo.basile@polito.it}
\author[3]{Marco~Torchiano}
\ead{marco.torchiano@polito.it}
\affiliation[1]
{
organization={Dipartimento di Ingegneria Elettrica e Elettronica, Universit\`a di Cagliari},
addressline={Via Marengo 3},
postcode={09123},
city={Cagliari},
country={Italy}
}
\affiliation[2]
{
organization={Istituto di Matematica Applicata e Tecnologie Informatiche "E. Magenes" (IMATI), Consiglio Nazionale delle Ricerche},
addressline={Via de Marini, 6},
postcode={16149},
city={Genova},
country={Italy}
}
\affiliation[3]
{
organization={Dipartimento di Automatica e Informatica, Politecnico di Torino},
addressline={Corso Duca degli Abruzzi, 24},
postcode={10129},
city={Torino},
country={Italy}
}
\cortext[cor1]{Corresponding author}

%\maketitle

\begin{abstract}

Evaluating the effectiveness of software protection is crucial for selecting the most effective methods to safeguard assets within software applications. Obfuscation involves techniques that deliberately modify software to make it more challenging to understand and reverse-engineer, while maintaining its original functionality. Although obfuscation is widely adopted, its effectiveness remains largely unexplored and unthoroughly evaluated. 
This paper presents a controlled experiment involving Master's students performing code comprehension tasks on applications hardened with obfuscation.
\changeda{The experiment's goals are} to assess the effectiveness of obfuscation in delaying code comprehension by attackers and to determine whether complexity metrics can accurately predict the impact of these protections on \changeda{success rates and durations of code comprehension tasks}.  
The study is the first to evaluate the effect of layering multiple obfuscation techniques on a single piece of protected code. It also provides experimental evidence of the correlation between objective metrics of the attacked code and the likelihood of a successful attack, bridging the gap between objective and subjective approaches to estimating potency. 
%Furthermore, the paper evaluates the strategies and processes employed by the subjects to approach obfuscated applications. 
Finally, the paper highlights significant aspects that warrant additional analysis and opens new avenues for further experiments.
\end{abstract}



\end{frontmatter}


%%%%%%%%%%%%%%%%%%%%%%%%%%%%%%%%%%%%%%%%%%%%%%%%%%%%%%
\section{Introduction}
\label{sec:intro}

Software is constantly threatened by various attackers who seek to misuse applications, steal intellectual property, or use software as a \changeda{vector} for more extensive attacks, such as malware infections. This attack scenario is referred to as a ``\mate attack''~\cite{collberg-falcarin}. \mate attacks can be carried out on the attacker's machine, where an array of tools is available to reverse engineer and tamper with the target applications. These tools include static, dynamic, symbolic, and concolic analysis tools, as well as debuggers, disassemblers, and decompilers, among others.

Hence, to protect their assets, software developers must rely on software protection techniques deployed directly into the software application and on remote, trusted servers under their control.
Among these techniques, obfuscation plays a major role. 
Obfuscation encompasses a family of techniques that intentionally alter \changeda{the target application's code} to make it more difficult to understand, analyse, or reverse-engineer, while preserving the application's functionality~\cite{collberg1997taxonomy}. \changeda{Although obfuscation cannot, in theory, prevent \mate attacks~\cite{barak2001pop}, in practice it is highly effective at delaying adversaries~\cite{viticchie2016assessment}. This protection significantly increases the difficulty and cost of reverse-engineering code, thereby acting as a deterrent. In real-world scenarios, an attack may be deemed unsuccessful when goals are not achieved within a reasonable timeframe, either because the expected reward no longer justifies the extended effort or because a new software version is released, forcing attackers to (almost) start over~\cite{basile2023riskanalysis}.}

Commercial and open-source tools are available to automatically apply obfuscation techniques, protecting target applications. However, the entire software protection field faces severe challenges. Commercial software protection tools and consulting services are expensive and opaque, relying heavily on \textit{security-through-obscurity}. There is no generally accepted method for evaluating the effectiveness of software protection~\cite{basile2023riskanalysis}.

In this context, determining the impact of obfuscation is a crucial research topic \changeda{that we address in this article.} Collberg introduced the abstract concept of \changeda{\textit{potency}} ~\cite{collberg1997taxonomy}. It refers to the degree to which an obfuscation technique increases the difficulty for humans of understanding, analysing, or reverse-engineering the obfuscated code. Hence, it measures how effectively obfuscation makes code harder for attackers to interpret. Potency is difficult to measure quantitatively, as it refers to human abilities in reverse-engineering the obfuscated code.

Collberg suggested, in the same paper, using quantitative complexity metrics computed before and after obfuscation to estimate potency \textit{objectively}. 
\sloc, \CC, size and complexity of the \cfg, \changeda{and} Halstead complexity measures are examples of the metrics used to measure potency~\cite{ceccato2014metrics}. Being introduced to estimate the software developer's difficulties in maintaining code,  they have never been related to attacker tasks in past research. 
Although it is reasonable to assume that code that is more complex for developers to maintain will be harder for attackers to understand, scientific evidence is still lacking.

Another alternative, \textit{subjective} approaches, consists of conducting empirical studies where different groups of subjects attempt to deobfuscate code \cite{icpc2017}. The analysis of success rates and time taken has been used to assess potency~\cite{viticchie20splitting}. The results of these experiments, conducted using state-of-the-art empirical methods, provide precise insights into obfuscation effectiveness. However, they are difficult to generalise to contexts significantly different from the objects of the experiments (\eg, larger attacked applications, different protections). Hence, more experiments with more subjects, other objects, and research objectives would be needed to gain sufficient insight to develop a precise and general definition of potency based on subjective approaches.

This article falls in the second category. 
We present the results of a controlled empirical experiment assessing the effectiveness of obfuscation techniques in hindering code comprehension by attackers aiming to compromise assets in software applications. 
%\textcolor{brown}{Code comprehension is a fundamental activity when attacking programs in \mate scenarios~\cite{collberg1997taxonomy,icpc2017}, since attackers typically need to understand the programs' business logic, control flow, and data dependencies to pinpoint the specific code that must be tampered with to port a successful attack.}
The experiment involved 152 subjects, Master's students in Computer Science Engineering from Politecnico di Torino (see Section~\ref{sec:subjects}). 
\changeda{The subjects performed tasks designed to assess the difficulty of comprehending code obfuscated with two variants of \cff implemented by Tigress\footnote{\url{https://tigress.wtf/}}, a FOSS obfuscator for the C programming language developed at the University of Arizona. One variant employs \ops to obfuscate the dispatch variable. \cff \cite{wangFlatteningTechReport} rearranges the basic blocks of the code, so that an attacker cannot readily obtain useful insights on the application's business logic by analysing its \cfg. \ops \cite{collberg1997taxonomy} are tautological Boolean expressions used to introduce bogus code that is never executed at runtime, thus increasing the amount of code that an attacker must analyse to understand the application's business logic. A complete description of \cff and \ops is available as supplementary material.}

% \textcolor{brown}{The controlled experiment included different phases. 
% The day before the experiment, we collected information about the subjects and estimated their programming ability, asking them to undertake a C programming test. 
% On the day of the experiment, subjects answered additional C programming questions, performed a warm-up task, and completed two main tasks. The tasks consisted of identifying and fixing bugs in three different software applications, the objects of the experiment (see Section~\ref{sec:objects}). The tasks aimed to assess the difficulty of comprehending the code and locating the bugs. Indeed, code comprehension is an activity expected to be hindered by obfuscation. Conversely, the bug corrections were straightforward on purpose (see Section~\ref{sec:attack_task}) \changed{and designed only to verify that successful participants actually understood the analysed code}, since our objective was to assess the impact of obfuscation on code comprehension. We wanted to prevent participants from spending significant time on the actual fixes rather than on understanding and locating the bugs. }

The research questions of the experiment aimed to evaluate the impact of two obfuscation transformations (see Section~\ref{sec:treatments}) on success rates and time. %\textcolor{brown}{The techniques used are two variants of \cff, one that splits the basic blocks to increase the complexity of the \cfg and another that uses \ops to increase the complexity of the \cff dispatch variable, the first two treatments in this experiment.}
Furthermore, the experiments sought to highlight the impact of layered protection, \ie, when both techniques are used together to protect the applications.
%\textcolor{brown}{, which is the third treatment. Layering multiple obfuscation techniques is a common practice in software protection and is supported by several code obfuscators, such as Tigress and Obfuscator-LLVM. These tools enable multiple transformations to be applied to the same code, aiming to compound the difficulty of reverse engineering. However,} 
To the best of our knowledge, no prior empirical study has quantitatively evaluated how combining obfuscation techniques affects the success rate or the time required to perform \changed{code comprehension} tasks. Trying to bridge the gap between objective and subjective approaches to estimate the potency, the experiment attempted to answer two additional research questions, \ie, whether complexity metrics can be used to accurately predict the impact of the treatments on success rates and attack time. 

This study has two major novelties.
\begin{enumerate}
\item To the best of our knowledge, it is the first study to assess the impact of layering multiple obfuscation techniques on the same protected code.
\item Furthermore, we provide experimental proof of the correlation between objective metrics of attacked code and the likelihood of a successful attack.
\end{enumerate}

For reproducibility purposes, we provide all questionnaires administered to the experiment subjects, their answers, the C source code that constitutes the treatments, the scripts used to analyse the gathered data, and the results obtained. All this information is contained in a replication package available on GitHub\footnote{\url{https://github.com/daniele-canavese/empirical-obfuscations}}.

This paper is organised as follows.
Section~\ref{sec:background} introduces the background and related work.
Section~\ref{sec:design} presents the design of the experiment, the research questions, the experimental procedure, and the analysis method.
Section~\ref{sec:results} presents the results of the analysis of the data collected with the experiment.
Section~\ref{sec:discussion} discusses the results we obtained.
Finally, Section~\ref{sec:conclusions} draws conclusions and outlines future work. 

%%%%%%%%%%%%%%%%%%%%%%%%%%%%%%%%%%%%%%%%%%%%%%%%%%%%%%
\section{Background and related works}
\label{sec:background}

\changeda{Software obfuscation is a family of protection techniques that aim to harden code against reverse engineering, by transforming the application's code to hinder human comprehension while preserving the original application's business logic.} Perfect obfuscation has been proven theoretically impossible by Barak \etal~\cite{barak2001pop}. 
Furthermore, some works have shown that \changeda{obfuscated code} can be recognised~\cite{dallaPredaOpaque}; hence, these techniques are not stealthy.
To the best of the authors' knowledge, no studies in the literature, either theoretically or empirically, demonstrate the effectiveness of these techniques against the reverse engineering of C programs; however, some works exist for Java~\cite{ceccato2009effectiveness}. \changed{Even worse, there is no consensus even on which aspects to consider for evaluating software protection~\cite{de2023evaluation}.}
Nonetheless, they have been extensively researched and employed for a long time in the software security industry to protect commercial software, \changeda{albeit reportedly uncommon in practice}~\cite{SoK2024}. It is implicitly assumed that they effectively delay reverse engineering tasks~\cite{cloakware}.

\changeda{Recent work has applied \ai to automate reverse engineering tasks~\cite{11098869}, and in particular, on both sides of the obfuscation/reverse engineering arms race. For example, \llms can produce effective obfuscations at the assembly level (dead‐code insertion, register substitution) that hinder analyst comprehension~\cite{aaai2025}, but may also be leveraged for automatic deobfuscation~\cite{PATSAKIS2024124912}. AI-based solutions have also been proposed to detect obfuscated code~\cite{10636038,CONTI2022103311}.}

%%%%%%%%%%%%%%%%%%%%%%%%%%%%%%%
\subsection{Layered protection}
\label{sec:background:layered}

Layered protection is a principle in software protection that suggests applying more than one protection technique to the same piece of code to further increase the effort required for an attacker to carry out their understanding tasks \cite{xu_layered_2020}. To be effective, the protections to apply need to reinforce each other. An example of synergistic techniques is adding anti-tampering techniques (\eg software attestation \cite{viticchie2016reactive}) to code protected with obfuscation. Obfuscation is expected to make understanding more difficult, and anti-tampering techniques make writing patches more difficult, as they need to be stealthy to evade the used anti-tampering techniques. 

As another example, Udupa \etal suggested using \ops to \changed{harden the understanding of the \cff dispatch variable behaviour \cite{udupaDeobf}}.
We deemed this a good example of layered protection, and therefore used it to generate one of the experimental treatments, as described in Section~\ref{sec:treatments}. 
Luckily, Tigress supports methods that obfuscate the dispatch variable of the \cff with \ops. \changed{For a more thorough presentation of the impact of layered protection, we refer to recently published papers~\cite{basile2023riskanalysis,de2023evaluation}.}

\subsection{Measures of the potency}
\label{sec:potency}

In the literature, two main approaches are used to assess the effectiveness of software protection techniques: 
% \begin{enumerate*}
\begin{inparaenum}[1)]
    \item theoretical evaluation based on software metrics;
    \item empirical evaluation based on controlled experiments involving students or on case studies with professional hackers. 
% \end{enumerate*} 
\end{inparaenum}

\textbf{Assessment based on code metrics.}
One of the early examples of the metrics-based approach is represented by \emph{potency}.  Collberg~\etal introduced this concept as a metric to estimate the effectiveness of obfuscation transformations on programs \cite{collberg1997taxonomy}.
Relying on the potency definition, Anckaert~\etal presented a comparison of obfuscation techniques \cite{anckaert2007obfuscation}.
Linn~\etal considered the confusion factor, which estimates the number of binary instructions that a decompiler is unable to parse \cite{linn2003obfuscation}.
Goto~\etal\ proposed a method to quantitatively measure the complexity of obfuscated code based on compiler syntax analysis \cite{goto2000quantitative}.
Udupa~\etal\ estimated the complexity increase in obfuscated programs using data that can be extracted with static and dynamic analysis tools \cite{udupaDeobf}.
Visaggio~\etal\ instead used code entropy as a protection potency metric for obfuscated JavaScript code \cite{visaggio2013empirical}.
Recently, alternatives to fully theoretical metrics-based evaluations have been proposed. Canavese~\etal presented a method to estimate \emph{a priori}  the effectiveness of software obfuscation by means of artificial neural networks~\cite{canavese2017estimating}. Van den Broeck~\etal \cite{van2021obfuscated} proposed a set of software metrics to evaluate the potency of layered obfuscation techniques in protecting software against MATE attacks. Zhao\etal~\cite{ZHAO2020102072} employed multiple code metrics to train a set of \changeda{deep neural networks} able to identify the obfuscation techniques used to protect a binary, to ease its deobfuscation. Raubitzek\etal~\cite{RAUBITZEK2024103850} evaluated the impact of layered obfuscation techniques on code metrics for malware classification purposes. 
To the best of our knowledge, the most investigated software protection technique in the literature is obfuscation, while other protection techniques (such as \changeda{\bs}) have not been assessed using objective approaches.

\textbf{Assessment based on controlled experiments with students.
}
Evaluations using experiments with human subjects were introduced by Sutherland \etal, who presented the first study in this field~\cite{sutherland2006empirical}. The authors correlated the expertise of attackers with the correctness of reverse engineering tasks. Moreover, they demonstrated that source code metrics are inadequate for estimating delays in attack tasks when binary code is involved.
Ceccato \etal measured, in two controlled experiments, the correctness and effectiveness in understanding and modifying decompiled obfuscated Java code, compared to decompiled clear code~\cite{ceccato2009effectiveness}.
This work was extended with a larger set of experiments on several obfuscation techniques in two successive works~\cite{ceccato2014family,ceccato2015large}. Their replication package was then used by H{\"a}nsch \etal to conduct a similar experiment and assess a slightly different set of obfuscations~\cite{hansch2018programming}.

Viticchi\'e \etal empirically evaluated with students the attack delay introduced by a data obfuscation technique, namely \textit{variable merge}, \changeda{reporting that attacks are delayed by a factor of six when that technique is used~\cite{viticchie2016assessment}.} Ceccato~\etal experimented with Master's students performing attack tasks on an application protected with \bs, which moves relevant slices of code to a trusted server, \changeda{highlighting an effective reduction in attack success rate and an increase in attack time}~\cite{viticchie20splitting}.

\textbf{Assessment based on experiments with professional participants.} Involving professional hackers creates an experimental context that closely resembles a real attack scenario. \changeda{Unfortunately, they are rare and costly.
Also, professionals often do not want to participate in controlled environments or be surveilled at work, leading to a lack of precise measurements and limiting the ability to observe the phenomenon in detail.}

Ceccato \etal conducted one of the first empirical evaluations of protection techniques involving professional hackers~\cite{icpc2017}, \changeda{asking them to attack fully protected applications} (\ie multiple protections applied to mimic real-world scenarios) to assess how professional attackers perceive and approach these protections. 
In a subsequent study, the same authors launched a public challenge with a cash bounty to validate the findings from the initial experiment in a broader context with professional white-hat hackers~\cite{emse2018}.
The papers present methods for coping with these limitations using ad hoc empirical settings.

\textbf{Open issues.}
A recent survey highlighted the lack of strength measurements for software protections~\cite{de2023evaluation}, arguing that authors of obfuscation tools might consider such measurements irrelevant, too hard or time-consuming to measure. 
Indeed, there is no consensus on which metrics to use in the objective field, and complexity metrics from the software engineering field may not be computable on obfuscated software, even with the most advanced disassemblers. In the subjective field, the number of experiments required to obtain a wide coverage of applications and protections is probably too large to be affordable. Moreover, some binary techniques can completely prevent the use of certain analysis tools, rendering complex comparisons of alternative attack strategies impossible. 
Hence, firms developing software protections resort to pen-testing. \changed{Furthermore, a framework has been proposed to evaluate the effectiveness of obfuscation techniques~\cite{bjorn-o-filosof}. Despite being in an early stage and mainly theoretical, it looks promising. The results of our experiments can be used to complement and confirm the predictability features.}

%%%%%%%%%%%%%%%%%%%%%%%%%%%%%%%%%%%%%%%%%%%%%%%%%%%%%%
\section{Design of the experiment}
\label{sec:design}
\subsection{Goals and Research Questions}
\label{sec:research_questions}

The existing literature reports evidence suggesting that obfuscation delays attackers when they have to perform understanding tasks~\cite{viticchie20splitting,ceccato2014family}. The assumption is that obfuscation reduces the likelihood that attackers can understand the code's semantics and mount attacks.
This hypothesis is also anecdotally confirmed by its extensive use in companies that provide software protection services~\cite{basile2023riskanalysis}.

The \emph{purpose} of this study is to evaluate how different obfuscation techniques can delay or prevent attackers from \changed{correctly understanding the code they want to tamper with}.
The \emph{quality focus} is the ability of the technique to reduce the attackers' capability to mount a successful attack within a useful time frame, by making the understanding phases more complex due to an increased perception of code complexity.
The study evaluates the \emph{perspective} of the attackers. In the experiment described in this paper, the role of the attacker is played by students with a consolidated minimum level of expertise in manipulating application source code. The above purpose can be achieved using an experiment aimed at answering the following research questions:

\begin{itemize}

	\item \textbf{RQ1 (treatment-success)}: Do specific obfuscation techniques have a different impact on the possibility for attackers to succeed in \changed{understanding an application's code} in a given time frame?

	\item \textbf{RQ2 (treatment-time)}: Do specific obfuscation techniques differently impact the time needed for attackers to succeed in \changed{understanding an application's code}?

	\item \textbf{RQ3 (complexity-success)}: Does the complexity of an application (measured with objective metrics) allow predicting the possibility for attackers to succeed in \changed{understanding an application's code} with the application in a given time frame?
	
	\item \textbf{RQ4 (complexity-time)}: Does the complexity of an application (measured with objective metrics) allow predicting the time needed for attackers to succeed in \changed{understanding an application's code}?

\end{itemize}

The answers to the above questions are interesting from both theoretical and practical perspectives. Theoretically, they can help specify a more precise formulation of the potency, which was left unspecified in the seminal work by Collberg\etal~\cite{collberg1997taxonomy}.  
Practically, this study will help software developers decide how to protect their applications and what expectations can be reasonably met by adopting obfuscation. 

We know from previous studies that obfuscation impacts, to various extents, attackers' activities \cite{viticchie20splitting,ceccato2014family,emse2018}. 
Hence, the first research question concerns the ability of the considered obfuscation techniques to sufficiently hinder comprehension of the protected code, thereby preventing an attacker from mounting a successful attack. In our experimental setting, it means whether, given a certain amount of time, the attack allows the goals to be reached.

In practical terms, we cannot expect protections to prevent any attacker from acting indefinitely. At best, they constitute a hurdle that slows them down. The defenders' purpose is to slow them down so that the attack is not economically viable. Thus, our second research question addresses the techniques' ability to increase the time required to successfully understand the target code in order to mount a successful attack.

The essential idea behind obfuscation techniques is to make understanding a program more difficult. %\textcolor{brown}{Hence, several obfuscation techniques aim at increasing the code's cognitive complexity. While the construct of code cognitive complexity is intuitively easy to grasp, many measures attempt to convert it into pragmatic terms.}
The third research question aims to investigate whether a cognitive complexity measure can be used to predict the success rate of attacks.

Analogously, the fourth research question focuses on whether code complexity can predict the time required to successfully understand the target code, thereby facilitating a successful attack. A positive answer to either of the last two questions could indicate that it is possible to build models that accurately estimate the effectiveness of obfuscation techniques based on code cognitive complexity.

%%%%%%%%%%%%%%%%%%%%%%%%%%%%%%%%%%%%%%%%%%%%%%%%%%%%%%%%
\subsection{Threat Model}
\label{sec:threat_model}

\changeda{The threat model for this empirical investigation is \mate, which is the reference threat model in this field and has already been used in another empirical study~\cite{viticchie20splitting} and a recent work presenting software protection as a risk analysis process~\cite{basile2023riskanalysis}.}

Indeed, in our experiment, the attacker has full access to the source code of a software application that can be executed on a local computer without requiring any server interaction.
In the real world, attackers mainly have access to binary code.
However, we assume that source code, almost as comprehensible as the original source, can be obtained with sufficient effort.
Experienced attackers can reconstruct high-quality representations of the original source code thanks to commercial or open-source tools\footnote{Examples of excellent reverse engineering and exploitation tools are IDA-Pro \url{https://www.hex-rays.com/products/ida/}, radare2 \url{https://rada.re/}, and Ghidra~\url{https://ghidra-sre.org/}}. 
Since the goal of this research is to evaluate the level of protection that can be achieved with selected obfuscation techniques, we are not interested in assessing \changeda{the additional delay due to binary code decompilation or disassembling} (although some obfuscation techniques can make these tasks harder). %\textcolor{brown}{It is worth noting that the fact that a \mate attacker can perform several complex operations does not imply that all of these will actually be worthwhile in the experiments presented in this paper.}

%%%%%%%%%%%%%%%%%%%%%%%%%%%%%%%%%%%%%%%%%%%%%%%%%%%%%%%%%%%%%%%%%

\subsection{\changed{Objects}}
\label{sec:objects}

The objects of this experiment are three open-source applications:

\begin{itemize}
    \item \lstinline|arithmetic|\footnote{\url{https://manpages.debian.org/stretch/bsdgames/arithmetic.6.en.html}}, taken from the \lstinline|bsdgames| Debian package: a quiz game asking the user to evaluate some simple mathematical computations;
    \item \lstinline|number|\footnote{\url{https://manpages.debian.org/bullseye/bsdgames/number.6.en.html}}, another application in the \lstinline|bsdgames| package, a program that converts Arabic numerals to English (\eg `42' becomes `forty two');
    \item \lstinline|tictactoe|, the listing 44 in the book C: The Complete Reference, 4th Ed. \cite{c_complete_reference}, an implementation of the tic-tac-toe game where the user can play against the computer.
\end{itemize}

Practical and methodological considerations guided the selection of these three applications. First, all programs are open-source and implemented in ANSI C. \changeda{Thus, compiling the source files does not require any specific compiler or flags.} The objects can be built on Linux, macOS, and Windows and were tested with the GCC, MSVC, and Clang compilers. The applications can be launched by running the built executable, \changeda{which presents an interactive command-line interface.} 

Second, they differ in functionality, covering arithmetic quizzes, string formatting, and interactive gameplay, thus providing some variation in application behaviour. Third, as reported in Table~\ref{tab:metrics}, the applications present increasing values of code complexity metrics and \sloc. Moreover, applications needed to remain small enough to ensure that participants could realistically complete the assigned tasks within the allotted time. 

Each program is accompanied by a README file that describes the program and provides instructions on how to build and launch it, without indicating whether the version is protected or not.

%%%%%%%%%%%%%%%%%%%%%%%%%%%%%%%%%%%%%%%%%%%%%%%%%%%%%%%%%%%%%%%%%

\subsection{Treatments}
\label{sec:treatments}

The treatments in this experiment are the protections applied to the programs (objects) presented in Section~\ref{sec:objects}.

\begin{itemize}
    \item The first treatment (\VersionCFF) consists of transforming the code using the \cff obfuscation with the \emph{split basic block} Tigress option. This option further divides the original basic blocks to increase the number of \lstinline|case| branches.
    
    \item The second treatment (\VersionCFFOp) consists of transforming the code using the \cff technique with the \emph{Flatten Obfuscate Next} Tigress option described in Section~\ref{sec:background:layered}. This option uses \ops to obfuscate the code that updates the CFF dispatch variable. 
    
    \item The third treatment (\VersionDoubleCFF) is a layered approach that transforms the code using \cff with both the \emph{split basic block} and \emph{Flatten Obfuscate Next} options.
\end{itemize}

Moreover, all the treatments were also invoked with the Tigress \textit{encode literals} obfuscation\footnote{This transformation replaces literal strings with calls to a function generating such strings at runtime. Details are available on the Tigress website at \url{https://tigress.wtf/encodeLiterals.html}. We obfuscated the functions to encode the literals with \cff to hinder a reconstruction attempt, and moved them into a separate file (\texttt{extra.c}), explicitly telling the students to ignore this file to avoid wasting their time.}, which removes the variables semantics and avoids simple string searches. The Tigress invocation parameters used to obfuscate the literals and the encoder function are reported \changeda{in the supplementary material.}

All the protected treatments exhibit an increase in the values of complexity metrics compared to their vanilla counterparts. The versions protected with a single option have similar complexity values, whereas, as expected, the applications protected with the double \cff show significantly higher metrics values. Table~\ref{tab:metrics} reports complexity metrics values computed with frama-c\footnote{\url{https://frama-c.com/}}. % on the three vanilla applications and their twelve protected variants. 
Since the three complexity measures strongly correlate with $R^2 > 99.5\%$, we will use only one of them to represent the application complexity. We select \sloc since it is the most intuitive and easiest to compute.

\begin{table}[tb]
    \centering
    % \adjustbox{\textwidth}{
    \begin{tabular}{lrrrr}
        \toprule
        \textsc{application} & \textsc{protection} & \textsc{SLOC} & \textsc{branches} & \textsc{CC}\\
        \midrule
        \lstinline|arithmetic| &    vanilla &  219 &  33 &  40\\
        \lstinline|arithmetic| &       \VersionCFF &  426 &  96 & 103\\
        \lstinline|arithmetic| &     \VersionCFFOp &  361 &  73 &  80\\
        \lstinline|arithmetic| & \VersionDoubleCFF & 1024 & 261 & 268\\
        \cmidrule(lr){1-5}
        \lstinline|number|     &    vanilla &  274 &  52 &  57\\
        \lstinline|number|     &       \VersionCFF &  373 &  78 &  83\\
        \lstinline|number|     &     \VersionCFFOp &  349 &  70 &  75\\
        \lstinline|number|     & \VersionDoubleCFF &  629 & 148 & 153\\
        \cmidrule(lr){1-5}
        \lstinline|tictactoe|  &    vanilla &  162 &  27 &  33\\
        \lstinline|tictactoe|  &       \VersionCFF &  259 &  56 &  62\\
        \lstinline|tictactoe|  &     \VersionCFFOp &  255 &  54 &  60\\
        \lstinline|tictactoe|  & \VersionDoubleCFF &  586 & 144 & 150\\
        \bottomrule
    \end{tabular}
    \caption{Complexity metrics of the vanilla and protected applications (SLOC = Source Lines of Code, CC = Cyclomatic Complexity).}
    \label{tab:metrics}
\end{table}

\subsection{Task}
\label{sec:attack_task}

We manually altered the source code of the three original applications to introduce a bug that was easily detectable while executing the programs and straightforward to fix once located in the code.
The students' task was to find and fix the bug. For the sake of readability, the code snippets in this section contain non-obfuscated literals. We introduced the following errors:

\begin{itemize}
    \item (\AppA) \textit{arithmetic}: we introduced a bit-wise AND between the variable \lstinline|result| holding the user's answer and a constant, so that the modified application prints a ``Wrong result'' message also when the user's answer is correct -- the task goal is to remove this extraneous operation;

{
\centering
\hspace{-60pt}
\noindent\begin{minipage}{.8\textwidth}
\begin{lstlisting}[caption=\AppA buggy version,frame=tlrb,numbers=none]{Name}
if(atoi(p)==(result<@\textcolor{red}{\&0xF0CACC1A}@> ))
{
    printf("Right!\n");
    ++nright;
    break;
}
printf("Wrong!\n");
\end{lstlisting}
\end{minipage}
\hspace{-120pt}
\begin{minipage}{.8\textwidth}
\begin{lstlisting}[caption=\AppA fixed  version,frame=tlrb,numbers=none]{Name}
if(atoi(p)==(result))
{
    printf("Right!\n");
    ++nright;
    break;
}
printf("Wrong!\n");
\end{lstlisting}
\end{minipage}
}

    \item (\AppN) \textit{number}: we introduced a wrong reference in the table \lstinline|name1| associating numbers with their textual representation that resulted in the wrong print (\eg \lstinline{2} is translated as \lstinline{three}) -- the task is to remove the wrong reference;
    
{
\centering
\hspace{-60pt}
\noindent\begin{minipage}{.8\textwidth}
\begin{lstlisting}[caption=\AppN buggy version,frame=tlrb,numbers=none]{Name}
if (*p != '0') {
    rval = 1;
    printf("%s", name1[*p - <@\textcolor{red}{'/'}@>]);
}
\end{lstlisting}
\end{minipage}
\hspace{-120pt}
\begin{minipage}{.8\textwidth}
\begin{lstlisting}[caption=\AppN fixed  version,frame=tlrb,numbers=none]{Name}
if (*p != '0') {
    rval = 1;
    printf("%s", name1[*p - '0']);
}
\end{lstlisting}
\end{minipage}
}
    
    \item (\AppT) \textit{tictactoe}: we modified the logic of the function that checks after each move if the user or the computer won the game so that, if the user wins a game, the variable \lstinline|matrix| representing the play-board is modified to make the computer win -- the task is to remove the code that modifies the play-board.
    
{
\centering
\hspace{-60pt}
\noindent\begin{minipage}{.8\textwidth}
\begin{lstlisting}[caption=\AppT buggy version,frame=tlrb,numbers=none]{Name}
/* check rows */
for(i=0; i<3; i++)  
if(matrix[i][0]==matrix[i][1] &&
   matrix[i][0]==matrix[i][2] &&
   matrix[i][0]!=' ') return matrix[i][0]<@\textcolor{red}{='O'}@>;
   
/* check columns */
for(i=0; i<3; i++)  
if(matrix[0][i]==matrix[1][i] &&
   matrix[0][i]==matrix[2][i] &&
   matrix[0][i]!=' ') return matrix[0][i]<@\textcolor{red}{='O'}@>;

/* test diagonals */
if(matrix[0][0]==matrix[1][1] &&
 matrix[1][1]==matrix[2][2] &&
 matrix[0][0]!=' ')
   return matrix[0][0]<@\textcolor{red}{='O'}@>;

if(matrix[0][2]==matrix[1][1] &&
 matrix[1][1]==matrix[2][0] &&
 matrix[0][2]!=' ')
   return matrix[0][2]<@\textcolor{red}{='O'}@>;
\end{lstlisting}
\end{minipage}
\hspace{-120pt}
\begin{minipage}{.8\textwidth}
\begin{lstlisting}[caption=\AppT fixed  version,frame=tlrb,numbers=none]{Name}
/* check rows */
for(i=0; i<3; i++)  
if(matrix[i][0]==matrix[i][1] &&
   matrix[i][0]==matrix[i][2] &&
   matrix[i][0]!=' ') return matrix[i][0];

/* check columns */
for(i=0; i<3; i++)  
if(matrix[0][i]==matrix[1][i] &&
   matrix[0][i]==matrix[2][i] &&
   matrix[0][i]!=' ') return matrix[0][i];

/* test diagonals */
if(matrix[0][0]==matrix[1][1] &&
 matrix[1][1]==matrix[2][2] &&
 matrix[0][0]!=' ')
   return matrix[0][0];

if(matrix[0][2]==matrix[1][1] &&
 matrix[1][1]==matrix[2][0] &&
 matrix[0][2]!=' ')
   return matrix[0][2];
\end{lstlisting}
\end{minipage}
}

\end{itemize}

The tasks involved editing the source code to restore the original behaviour.
Such modifications are trivial. The most complex part of the tasks was understanding the source code to identify the locations to change. 
The tasks were designed to assess the difficulty in comprehending the protected code, which represents the area where obfuscation techniques are applied, without requiring significant effort during the tampering phase.%\textcolor{brown}{While fixing bugs is not an attack activity \textit{per se}, in practice, it is by no means different from cracking software since both require an intensive code comprehension phase.}

%%%%%%%%%%%%%%%%%%%%%%%%%%%%%%%%%%%%%%%%%%%%%%%%%%%%%%%%%%%%%%%%%%%%
\subsection{Subjects}
\label{sec:subjects}

The participants in the study are 152 Master's students in Computer Science Engineering at Politecnico di Torino, selected from those attending the ``Computer Systems Security''\footnote{The course covers both theoretical and practical levels of all the basics of ICT security and risk analysis, cryptography, authentication systems, X.509, PKIs and e-documents, security of IP networks and network applications, firewall and IDS/IPS, email security \url{https://security.polito.it/~lioy/02krq/}} 
course, held in the second (and final) year of the degree.
\changeda{They possess good skills in C programming, thanks to the degree's mandatory courses.} Therefore, they can be considered valid candidates to play the role of attackers of the object applications, as in past empirical studies~\cite{emse2018}. 
However, subjects were not expected to have any knowledge about \mate scenarios, attacks, and attack strategies. Indeed, the students involved had not attended any courses on software tampering or software reverse engineering. While some students may have acquired knowledge on such topics through informal means, our primary concern was to ensure that all participants possessed the minimum programming skills necessary to perform the assigned tasks. Since our goal was not to study the impact of expertise on performance, we did not attempt to measure or control for advanced background in software security. Moreover, subjects were screened based on their motivation, as participation was voluntary.
Participants who performed diligently on all tasks (regardless of success) were awarded 2 extra points out of 30 on the final course exam.

\begin{figure}[ht]
\centering
    \begin{subfigure}{0.49\linewidth}
        \centering
        \includegraphics[width=1\textwidth]{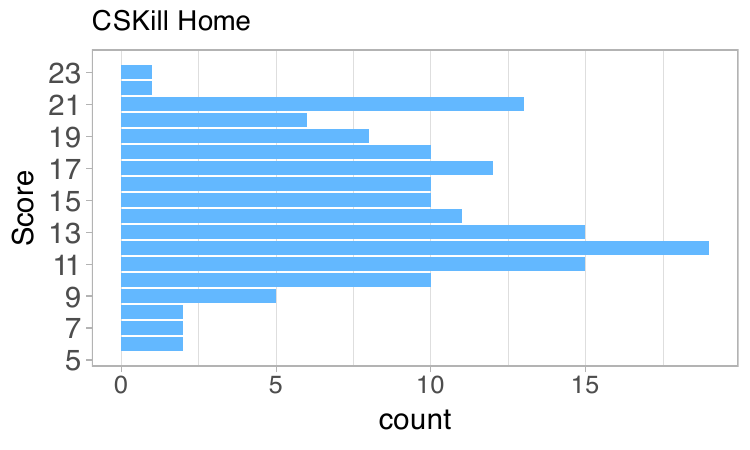}
        \caption{Home skill test results.}
        \label{fig:CtestresultHome}
    \end{subfigure}
    \begin{subfigure}{0.49\linewidth}
        \centering
        \includegraphics[width=1\textwidth]{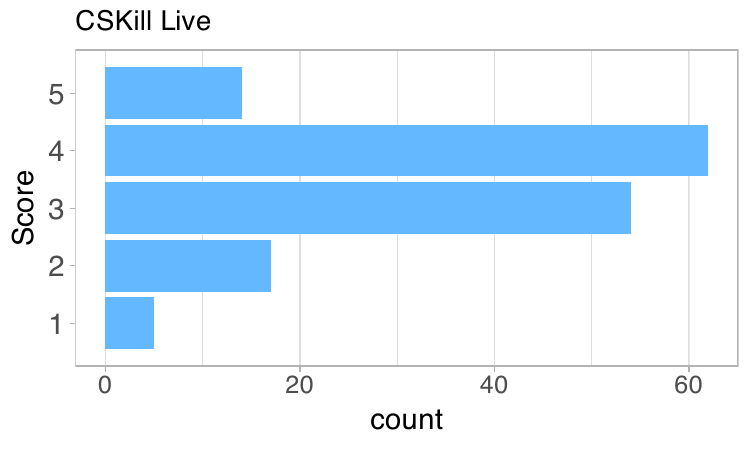}
        \caption{Live skill test results.}
        \label{fig:CtestresultLive}
    \end{subfigure}
    \caption{Score distribution of the C skill tests.}
    \label{fig:Ctestresult}
\end{figure}

Nonetheless, we evaluated the students' C skills before and during the experiment. 
In particular, before the experiment, we asked the participants to complete an online C test (\textit{home test}).
The test consisted of 23 multiple-choice questions selected from online resources\footnote{We selected questions from \url{https://www.proprofs.com/quiz-school/story.php?title=test-your-c-skills} and \url{http://www.pskills.org/c.jsp}}, \changeda{covering theoretical aspects of C programming (\eg pointers, static variables, standard C libraries) and compilers' behaviour (\eg calls to functions, variables initialisation, stack overflows). 
Most questions aimed to assess code comprehension abilities (\eg, asking the expected output of a small piece of code).}
\changeda{We opted for a multiple-choice test instead of a programming exercise because we were more interested in evaluating the students' abilities to comprehend and mentally simulate pieces of code, than in assessing their programming proficiency.}
Indeed, the task was a comprehension \changed{one} followed by a mostly trivial code change.
The overall results of the test are reported in  Fig.~\ref{fig:CtestresultHome}. 
Most students correctly answered at least half of the questions, thus demonstrating that they had sufficient background knowledge to participate in the experiment.

To confirm such results, before starting the experiment, we asked the subjects to answer five additional multiple-choice questions (\textit{live test}), whose results are summarised in Fig.~\ref{fig:CtestresultLive}. The students were able to solve most of the questions in this case.
%
%\textcolor{brown}{Then, we identified the lazy students (who achieved good results during the live test but poor scores at home), those who did not invest enough effort in answering the home test, and cheaters (who achieved excellent results at home but poor scores during the live test). Such students probably answered the home test with the help of a compiler and online resources.
%We have removed them and excluded their data from our analysis, which will be presented in Section~\ref{sec:results}. However, comparing the results with and without lazy students and cheaters, the analysis results remained unchanged.}

We also collected self-claimed information from subjects (\textit{self-data}) about:
\begin{itemize}
    \item \textit{their experience in C programming}, the great majority (78\%) of them have at least 1 year of programming experience in C;
    \item \textit{experience as part-time or full-time programmers}, about 20\% had experience as part-time or full-time professional programmers;
    \item \textit{their experience using an IDE environment for C programming} showed that when they programmed in C, they did it with an IDE. All the students reported that they had a good experience with at least one of the supported IDEs (Visual Studio, CodeBlocks, Xcode, Eclipse);

    \item \textit{their ability to use a debugger}, the great majority (87\%) were able to do all the most important operations (add breakpoints, execute the program step-wise, inspect the program variables);
    \item \textit{their experience with assembly programming}, which is at least 6 months for more than 80\% of the subjects;
    \item \textit{their experience in reverse engineering}, the great majority (89\%) had less than 3 months of experience.
\end{itemize}

%%%%%%%%%%%%%%%%%%%%%%%%%%%%%%%%%%%%%%%%%%%%%%%%%%%%%%%%%%%%%%

\subsection{Experimental procedure}
\label{sec:procedure}

The experimental procedure comprises two main phases:
\begin{enumerate}
	\item the preparation of the experiment, including the preliminary information gathering and the home skill test; in this phase, we have provided subjects with information needed to fill some of the post-experiment reports;
	\item the controlled experiment, which consisted of the live C programming test and self-data collection, followed by three \changed{code comprehension} tasks, one warm-up and the two main tasks.
\end{enumerate}
In preparing the experiment, we followed the same process as in past research in the empirical assessment of software protection~\cite{viticchie2016assessment,viticchie20splitting,emse2018}.

%%%%%%%%%%%%%%%%%%%%%%%%%%%%%%%%%%%%%%%%%%%%%%%%%%%%%%%%%%
%\label{sec:procedure:preparation}
\textbf{Preparation of the experiment.} After presenting the ``Hacking Experiment'' to the Computer Systems Security course class, students willing to participate had to send an email to pre-register.

We decided to let the subjects use their PCs to carry out the experiment, to have them execute the task in the most familiar setting, which is also closer to what happens to real hackers.
Thus, eleven days before the experiment, candidates were asked to perform a straightforward task to simulate the conditions of the controlled experiment, verify that they had the necessary background to conduct the experiment, and confirm that their environment was ready for the experiment. 
They received, by email, a per-user-ciphered zip file along with a password.
They had to open the archive, save the C source files, and report an integer value computed using their student ID number as input, forcing them to run a debugger, set a breakpoint, and read the value of a specific variable. After submitting the proper value, the candidates were officially accepted as subjects and received the link to the Google form with the home C programming test.

The subjects were invited to attend a one-hour seminar\footnote{The seminar was held nine days before the experiment in the time and room booked for the Computer Systems Security class during the rest of the semester.} introducing them to basic information about obfuscation techniques and attack tasks. \changeda{In particular, we presented an excerpt of a glossary\footnote{\url{https://drive.google.com/file/d/1RIuNkodHIL7QqLGrtNv5NI-Jzm-cZMq7/view?usp=sharing}} explaining a set of generic attack steps from the ontology published in a previous paper \cite{emse2018}, asking the subjects to read/study it before the experiment.}
These attack steps were needed to annotate the post-experiment reports, where participants described \changed{the procedure they followed when carrying out the tasks}.

%%%%%%%%%%%%%%%%%%%%%%%%%%%%%%%%%%%%%%%%%%%%%%%%%%%%%%%%%%%%%%%%
%\label{sec:procedure:experiment}
\textbf{Controlled experiment.} The day before the experiments, the subjects received an email with three numbered and encrypted zip files containing the three treatments assigned to them. The email did not include the passwords, which were only made available at the beginning of each task. The purpose was to send a reminder and avoid network overloads on the day of the experiment. The subjects were explicitly asked to follow a well-defined procedure during the experiment:
\begin{enumerate}
\item undertake the live C programming test;
\item answer the questionnaire to collect the self-data;
\item perform a warm-up task; and
\item perform the two main tasks. 
\end{enumerate}

The \textit{warm-up task}, lasting 40 minutes, asked the subjects to fix a bug in a vanilla application (see Section~\ref{sec:objects}).
This task aimed to avoid issues (e.g., incompatible environments and problems related to the network and the personal laptops) that could have affected the subsequent two phases of the experiment.
Moreover, we wanted all the subjects to be familiar with the intended task, which, as explained in  Section~\ref{sec:attack_task}, consisted of identifying the line of code to modify and applying a straightforward modification that corrected the bug.
%\textcolor{brown}{Indeed, we learned from past experiments \cite{viticchie2016assessment,viticchie20splitting} that despite our efforts in the preliminary phases, a non-negligible number of students either do not understand the task or have problems with their setup, making it hard to compare the first task, when a greater number of students has issues, with the following tasks, where the same issues are typically solved.}

At the beginning of this phase, participants received the password to access the first archive and a link to a Google form to submit their answers.
Either at the end of the time frame or before, if they were convinced that their task was successful, they were asked to fill out another Google form \changeda{with the following information: if they succeeded in executing the task; the amount of time in minutes that they needed; the OS, IDE and debugger they used; a qualitative evaluation of the task clarity and difficulty; a description of how they performed the task as a sequence of independent steps, each of them associated with a label (from the glossary), the target variable or function, and a textual description of the activity perfomed; an archive containing the application source file they fixed and a screenshot showing a successful run of the fixed application.}
% \begin{itemize}
%     \item the amount of time in minutes that they needed to execute the task;
%     \item  to perform the task;
%     \item if they succeeded in executing the task;
%     \item a qualitative evaluation of the task clarity and difficulty;
%     \item a description of \changed{how they performed the task} as a sequence of \changed{independent} steps; for each of the latter, they were asked to assign them a \changed{label} (from the glossary), the target variable or function, and a textual description of \changed{the activity perfomed}; 
%     \item an archive containing the application source file they fixed and a screenshot showing a successful run of the fixed application.
% \end{itemize}

The \textit{two main tasks}, lasting 70 minutes each, asked the subjects to fix the bug in obfuscated applications. 
Each participant received one of the two objects they did not receive in vanilla form during the warm-up; one was protected with \VersionCFF or \VersionCFFOp and one with \VersionDoubleCFF, not necessarily in this order. That is, some subjects received the version protected with the double options before a version with a single option enabled.  While the participants performed their tasks, the researchers answered all their questions, except those concerning how to carry out the task.
We note that, despite careful testing conducted on all platforms, we had to assist the participants in resolving a few setup issues. However, these were mostly limited to the warm-up phase.

After the experiment, the actual status of the task (successful or failed) was confirmed by the organisers using a standard assessment procedure:
\begin{enumerate}
    \item extract the fixed source file and the screenshot showing a successful run of the application from the submitted archive;
    \item automatically check the fix's correctness by compiling the modified source code and launching the obtained binary file against a collection of test cases. A manual effort and a collegial decision were devoted to addressing the false negatives, which were subjects who declared a failure but passed all the tests, and the false positives, which were subjects who declared they succeeded but indeed failed the tests.
    A task was deemed successful only if the source code compiled without errors and all test cases were passed. 
\end{enumerate}

% The question asked through the Google Forms, the subjects' answers to such forms, the results of the C skill questionnaires, and the time needed to execute the \changed{code comprehension} task are the data on which we base our empirical analysis, and are available in the replication package.

%%%%%%%%%%%%%%%%%%%%%%%%%%%%%%%%%%%%%%%%%%%%%%%%%%%%%%%%%%%%%%%%%
\subsection{Variables}
\label{sec:variables}

The research questions presented in Section \ref{sec:research_questions} above guided us in selecting the variables to collect during the experiment and formulating the hypotheses.

As \emph{dependent variables}, we consider the following aspects of the executed \changed{code comprehension} tasks:
	
\begin{description}
    \item[\textbf{Succeeded}:] corresponds to the participant's ability to complete successfully or not; it has been assessed as explained in Section~\ref{sec:procedure}.
    Variable $Succeeded(s_i,t_i)$ is a Boolean, true if the task $t_i$ by subject $s_i$ was successful.

    \item[\textbf{Time}:] represents the elapsed time to perform the task. The time is self-reported\footnote{We cross-checked the self-reported times against the submission timestamp of the Google forms used by the participants to submit their answers, without finding any inconsistencies.}
    Variable $Time{(s_i, t_i)}$ measures the minutes spent by subject $s_i$ to perform the task $t_i$, regardless of \changed{the participant success in performing the task.}

    We remark that we collected time for all experimental tasks. Because the failed attempts mostly lasted until the end of the allotted time, the analysis will only consider the time for the successful \changed{attempts} (\ie when $Succeeded == \texttt{true}$).
\end{description}

The \emph{independent variables} we consider are the following:
\begin{description}
\item[\textbf{Treatment}:] indicates the type of transformation applied to the source code: \{ \textit{Vanilla}, \VersionCFF , \VersionCFFOp , \VersionDoubleCFF \}, as described in Section~\ref{sec:treatments}. 
The \textit{Vanilla} treatment consists of no obfuscation and has been used in the warm-up task performed during period 1.

\item[\textbf{Application}:] indicates the application on which the task was performed. In our experiment, we had three applications: \{ \textit{arithmetic}, \textit{number} , \textit{tictactoe} \}.

\item[\textbf{Period}:] indicates the period in which a task was performed. Our experiment consisted of three periods, where period 1 served as a warm-up and was not considered in the hypothesis testing.

\item[\textbf{Sequence}:] indicates the specific succession of treatments administered to a given subject, which defines the experimental group the subject was assigned to. The hypothesis testing did not consider the information about the treatment used in period~1.

\item[\textbf{CSkill Home}:] indicates the C language skills of the subjects. The score is based on the aggregate results of the (pre-experiment) home test that the participants filled in from home. It ranges from 1 to 23.

\item[\textbf{CSkill Live}:] indicates the C language skills of the subjects. The score is based on the aggregate results of the live tests. It ranges between 0 and 5.

\item[\textbf{SLOC$_{App}$}:] indicates the size in LOC of the application that was targeted by the task -- \ie after obfuscation has been applied. As discussed in section \ref{sec:treatments}, this metric represents the complexity of the application.

\end{description}

\subsection{Hypotheses}
\label{sec:hypotheses}

Based on the four research questions, we formulate the following null hypotheses to be tested based on the variables defined above:
\begin{itemize}
    \item $H_{ss0}$: there is no difference in \changed{code comprehension} success rate between the two basic obfuscation transformations.
    \item $H_{sl0}$: the use of layered protections has no effect in terms of \changed{code comprehension} success rates with respect to single obfuscation transformation.
    \item $H_{ts0}$: there is no difference in \changed{code comprehension} time between the two basic obfuscation transformations.
    \item $H_{tl0}$: the use of layered protections has no effect on \changed{code comprehension} time with respect to single obfuscation transformation.
    \item $H_{ms0}$: there is no difference in prediction accuracy, as far as the \changed{code comprehension} success rate is concerned, if objective metrics are used together with the combination of application and obfuscation technique.
    \item $H_{mt0}$: there is no difference in prediction accuracy, as far as the \changed{code comprehension} time is concerned, if objective metrics are used together with the combination of application and obfuscation technique.
\end{itemize} 

%%%%%%%%%%%%%%%%%%%%%%%%%%%%%%%%%%%%%%%%%%%%%%%%%%%%%%%%%%%%%%%%%
\subsection{Analysis method}
\label{sec:analysis_method}
The experimental measures are first summarised with basic descriptive statistics. The \changeda{success} rate is reported in terms of both absolute numbers of successful \changed{attempts} and proportion. For \changed{time}, we report the mean and standard deviation.

We plan to test the first four hypotheses using frequentist hypothesis testing.
Since we adopted a factorial crossover design, as recommended by~\cite{vegas2015crossover}, we opted to analyse the data using repeated measures mixed models -- both logistic and linear -- that included the sequence and period design variables. This choice enabled us to address the potential threats to validity arising from the selected design. We analysed the variance of such a model to check the statistical significance of the factors. We decided not to use non-parametric tests because the sample size -- 152 participants for a total of 304 data points -- may be sufficient to satisfy the conditions of the central limit theorem~\cite {de2013using}; this condition allows us to interpret the results even in the presence of slight departures from normality. Moreover, mixed models have fewer constraints compared to other methods in the case of repeated measures ANOVA.
	
To test hypotheses about \textbf{Success} -- \ie $H_{ss0}$ and $H_{sl0}$ -- we use a within-subjects logistic mixed model regression, where the success rate is the dependent variable and the independent variables are the indicator variables for the levels of the variables Treatment, Application, Sequence, and Period. The reference levels (\ie the intercept) correspond to Treatment = \VersionCFF, Application = \textit{arithmetic}, Period=2, and Sequence = \VersionCFF $\rightarrow$ \VersionCFF + \VersionCFFOp.
The participant ID represents the random component of the model.
This model allows us to accurately assess the effect of the main factor in a within-subject crossover design. The first hypothesis ($H_{ss0}$) will be tested by looking at the \changed{statistical} significance of the coefficient $\beta_{\VersionCFFOpMATH}$ corresponding to the effect of adopting \VersionCFFOp instead of \VersionCFF. 
The second hypothesis ($H_{sl0}$) will be tested by looking at the \changed{statistical} significance of the coefficient $\beta_{\VersionDoubleCFFMATH}$ corresponding to the effect of adopting \VersionDoubleCFF.
In addition to the statistical significance, we will look into the effect size of the Treatment. The $\beta_\cdot$ coefficients represent the log-odds-ratio of the application of the treatment level, therefore $e^{\beta_\cdot}$ is the odds-ratio.
	
To test the hypotheses concerning \textbf{Time} -- \ie $H_{ts0}$ and $H_{tl0}$ --, we use a within-subjects logistic mixed model regression, where the dependent variable is Time, while the predictors are the same as the previous model.
We only considered the participants who successfully completed the task.
The hypothesis ($H_{ts0}$) -- concerning the difference between simple obfuscations -- will be tested by looking at the \changed{statistical} significance of the coefficient $\beta_{\VersionCFFOpMATH}$ corresponding to the effect of adopting \VersionCFFOp instead of \VersionCFF.
The hypothesis ($H_{tl0}$) -- related to the effect of layering two protections -- will be tested by looking at the \changed{statistical} significance of the coefficient $\beta_{\VersionDoubleCFFMATH}$ corresponding to the effect of adopting \VersionDoubleCFF.
In addition to the statistical significance, we will look into the effect size of the treatment. The $\beta_\cdot$ coefficients represent the average variation in Time due to the application of the treatment level.

As far as hypotheses concerning the \textbf{prediction accuracy} -- \ie $H_{ms0}$ and $H_{mt0}$ --, we build models analogous to the two described above by replacing the indicator variables for Treatment and Application with the size of the application (function).
To test $H_{ms0}$, we observe that the alternative models can be plotted using a \roc curve and the \auc computed. The decision on the hypothesis is based on the comparison of the \changed{AUCs} using DeLong's test for two correlated ROC curves~\cite{DeLong88}.
To test $H_{mt0}$, we compare the goodness of fit of the two alternative models using both AIC (Akaike information criterion) and BIC (Bayesian information criterion). Since the models are not nested, a Likelihood Ratio test is not possible.
The statistical test results are assessed assuming \changed{statistical} significance at a 95\% \changed{confidence level (significance level $\alpha$=0.05).}
Hence, we reject the null-hypotheses when $p\mathrm{-value}<\alpha$. 
	
All the data processing is performed with the R statistical package\footnote{\url{https://www.R-project.org/}}. In particular, the mixed model regression was conducted using the \texttt{lme4} and \texttt{lmerTest} packages~\cite{lme4,lmerTest}, the AUC test using the \texttt{pROC} package~\cite{pROC}. 

%%%%%%%%%%%%%%%%%%%%%%%%%%%%%%%%%%%%%%%%%%%%%%%%%%%%%%%%%%%%%%%%%
\subsection{Threats to validity}
\label{sec:threats_to_validity}

We verified the design of our experiment against the checklist of the threats to validity reported by Wohlin \etal \cite{wohlin12}, which concern construct, internal, external, and conclusion validity.

\textbf{Construct validity.}
First, we report the threats to the \textit{construct validity}, which deal with the relationship between the theoretical constructs and the actual metrics collected for the experiment.

We evaluated the success of a task as a Boolean outcome. 
While this is a straightforward metric, it reflects a real-case scenario where either the attacker reaches its goals or not within the time frame when assets have a value.

\changeda{The C programming expertise was assessed using two distinct tests. Both were
complex and comprehensive enough to accurately assess students' abilities.} Moreover, using C programming tests is standard practice for recruiting programmers used by several companies worldwide~\cite{hiringTests}. We assured the participants that the test results would only be used for experiment-related purposes. \changeda{The first test was conducted in an uncontrolled environment, whereas the second was performed in a controlled setting to cross-check the former test's results. In this way, we identified minor anomalies (\eg, students who cheated in the first test) and excluded them.} We also noticed that their inclusion would not have changed the results of our analysis.

We allowed the participants to use any tool they were familiar with to complete their task. On the one hand, this approach could represent a confounding factor influencing the results, as a better tool could help complete the task faster and more precisely. On the other hand, we deemed this approach less impactful than forcing the user to use a specific tool with which they could not be very familiar.
We tried to mitigate this threat by asking participants to report their strategy, which could reveal the role of tools; moreover, we evaluated the impact using process analysis (see~\ref{sec:process_mining}).

\textbf{Internal validity.}
\changeda{Then, we address threats to the \textit{internal validity}, which may affect the ability to capture a cause-and-effect relationship between the independent variables and the experimental outcomes.}
We must consider all the noise factors that may indirectly influence the outcomes and attempt to mitigate or measure such effects.

\changeda{Literature reports that professional hackers use quite sophisticated tools to perform both reverse engineering and comprehension tasks \cite{icpc2017,emse2018}, in particular, in the presence of protected software. Using such tools (\eg decompilers), they can obtain accurate representations of the code to attack and to reconstruct the source code (almost) completely.}
Since the participants in our experiment were provided with the source code, we can substantially rule out the need for these tools. Giving the source code is the worst-case scenario -- the best case from the attacker's point of view -- simulating that accurate reverse engineering tools can recover exactly the original source code.

Another possible threat is that participants may not be aiming for the correct objectives. \changeda{Right before the experimental activities, researchers explained the tasks and objectives to all participants. Post-experiment questionnaires reported no comprehension issues. Nonetheless, reports indicated that a few participants did not fully digest the provided material. We evaluated these minor misunderstandings and confirmed they did not affect the ability to perform the tasks.}
 
The experiment has been conducted in a single session, thus we can exclude all the common threats related to time and repetitions (\eg history, testing, mortality, and statistical regression among experiments). A maturation effect may have occurred during the experimental session because each subject was assigned three tasks in sequence. Isolating the first warm-up task was also intended to reduce the maturation effect. Moreover, we have evaluated the impact of maturation between the two main tasks.

Participants were randomly assigned to treatment groups. \changeda{Since they are all students in the same year and course, we assumed their backgrounds were homogeneous.} A posteriori, we verified this assumption; even if their C skills were not very similar, the distribution of students to treatments resulting from random assignment was balanced.

The number of subjects assigned to each task is not equal. 
\changeda{We assumed that the third treatment (layered protection) was the most complex. Thus, we have assigned more subjects to it, to avoid an insufficient number of successful tasks,} which would have reduced the relevance and the impact of our analysis.
We randomly sampled the data to have a uniform distribution, and the analysis confirms that the unbalanced distribution does not affect the results.

\changeda{We encouraged the students' participation by granting a 2/30 bonus\footnote{In the Italian University system, grades are assigned on a 30-value scale.} for the Computer System Security exam grade. We encouraged the participants to do their best by promising in advance that, if they completed all the assigned tasks diligently, they would receive the bonus, regardless of the task's success or the usefulness of the information provided in the report. In our opinion, and based on our experience with previous optional activities conducted with students \cite{viticchie20splitting}, offering no incentives at all was not feasible, as students typically do not enjoy spending time on non-profitable academic tasks. Assigning the bonus to all participants led to high participation, but introduced the risk of noise into the collected data, as some subjects might have been primarily motivated by the bonus. Therefore, we added several checks, both to identify subjects who did not score well on the C tests due to their expertise, and to assess the quality of reports, to identify subjects who did not properly document their activities.}

\textbf{External validity.}
The risks to \textit{external validity} could limit the applicability of the findings, particularly in cases where professional attackers attempt to undermine obfuscated real-world applications.

Professional hackers might be better candidates for assessing the exploitation of \mate attacks; however, involving them presents significant challenges \cite{basile2023riskanalysis}.
Although students' hacking program expertise falls short of that of professional hackers, we assumed that the problem-solving skills of top students are likely comparable to those of hackers. Given that expertise level impacts the Success variable -- \changeda{given the tight timeframe to complete tasks} -- we selected the subjects and treatments to ensure \changeda{sufficient successes within the allotted time}. Additionally, we recruited participants on a voluntary basis to highlight motivation levels, presuming that higher motivations would closely align with typical hacker profiles. Furthermore, we assessed the effectiveness of the obfuscation techniques by comparing subjects' performance on both vanilla and obfuscated application versions, thus demonstrating the influence of expertise on task performance. The ongoing discussion about using students as substitutes for professionals in empirical software engineering has been longstanding and has recently gained more attention~\cite{Feldt2018}. In controlled laboratory experiments designed to address research questions that require strict conditions, it may be justifiable to sacrifice some realism for a more controlled environment. Moreover, the distinction between the characteristics of students and professionals is not clear-cut and shows considerable overlap. Several studies involving students have recently been published~\cite{viticchie2016assessment,viticchie20splitting}.

We lack sufficient findings to assess how the studied techniques might safeguard programs that are significantly different (\eg larger or more complex) than the ones we have examined, even though we explore the connections with complexity metrics. Indeed, the objects of our experiment are relatively small compared to commercial programs that hackers frequently target. \changed{However, attackers typically spend significant time and resources to locate the code that needs to be fully comprehended to mount an attack~\cite{basile2023riskanalysis}. Large applications follow a modular structure; hence, once an attacker has identified the relevant parts, the code required to comprehend and perform their tasks successfully is comparable in size to our experiment's objects~\cite{emse2018}.
Locating the code to comprehend is a different mental activity, whose effort cannot be estimated with our experiment data.} 

Our study focuses on a specific obfuscation technique, \cff, hardened in one of the treatments by combining it with another obfuscation technique, \ops. Thus, the results presented in this paper may not be generalisable to other obfuscation techniques. Nevertheless, \cff is routinely employed to safeguard software assets, being supported by many code obfuscation frameworks (\eg Tigress\footnote{\url{https://tigress.wtf/flatten.html}}, LLVM Obfuscator\footnote{\url{https://github.com/obfuscator-llvm/obfuscator/wiki/Control-Flow-Flattening}}, JScrambler\footnote{\url{https://jscrambler.com/blog/jscrambler-101-control-flow-flattening}}, DashO\footnote{\url{https://www.preemptive.com/products/dasho/features/}}, Allatori\footnote{\url{https://allatori.com/features/flow-obfuscation.html}}). Furthermore, a recent survey~\cite{de2023evaluation} reports that \cff and \ops are among the most researched code obfuscation techniques in the state of the art.

\changeda{All participants are enrolled in the identical Master's program. This may lead to a bias in how subjects perform the tasks, linked to the specific educational curriculum implemented at Politecnico di Torino. Indeed, Master's students from different universities may have differing backgrounds in terms of programming languages, frameworks, styles, and methodologies, leading to varying performances when executing tasks.}

The study utilised three distinct programs. It is uncertain whether applications with varying structures or from different fields would produce comparable outcomes, even though the architectures identified in our applications are quite typical and present in numerous systems. Our evaluation clearly indicated that the targeted application influences \changed{code comprehension time}. Additional experiments are necessary to establish a connection between program structures, semantics, and the complexity of the tasks.
 
As mentioned regarding internal validity, the subjects utilised the application's source code to \changed{perform the tasks}. This situation does not reflect the scenario of attackers who only have access to the binaries. We recognise that \changed{understanding} binary code is more intricate than reverse engineering the source code; hence, we can view our situation as a worst-case scenario from the defender's perspective. Attackers may consider various strategies that do not involve an initial attempt to obtain more effective representations of the binaries (such as the source code). In future studies, we aim to assess the differences in attack duration and success rates when using binaries as targets.

\textbf{Conclusion validity.}
Finally, \textit{conclusion validity} threats concern the validity of the statistical methods used to derive outcomes from the data. We have used logistic and linear mixed models appropriate for the within-subject crossover design adopted in our experiment and used the suitable tests presented in Section~\ref{sec:analysis_method}. We have collected data using survey questionnaires designed according to standard methods and scales \cite{oppenheim92}, and used multiple-choice methods to assess the C skills of the subjects. Tasks were similar and balanced, and subjects were not heterogeneous, as they were all master students; hence, experiments avoided random irrelevance.

%%%%%%%%%%%%%%%%%%%%%%%%%%%%%%%%%%%%%%%%%%%%%%%%%%%%%%
\section{Results}
\label{sec:results}
In this section, we present the analysis results obtained from the data gathered during the experiment, along with the subsequent answers to the Research Questions outlined in Section~\ref{sec:research_questions}.

\subsection{RQ1 - Success rate}
\label{sec:results:success}

\begin{figure}
    \centering
    \includegraphics[width=0.8\textwidth]{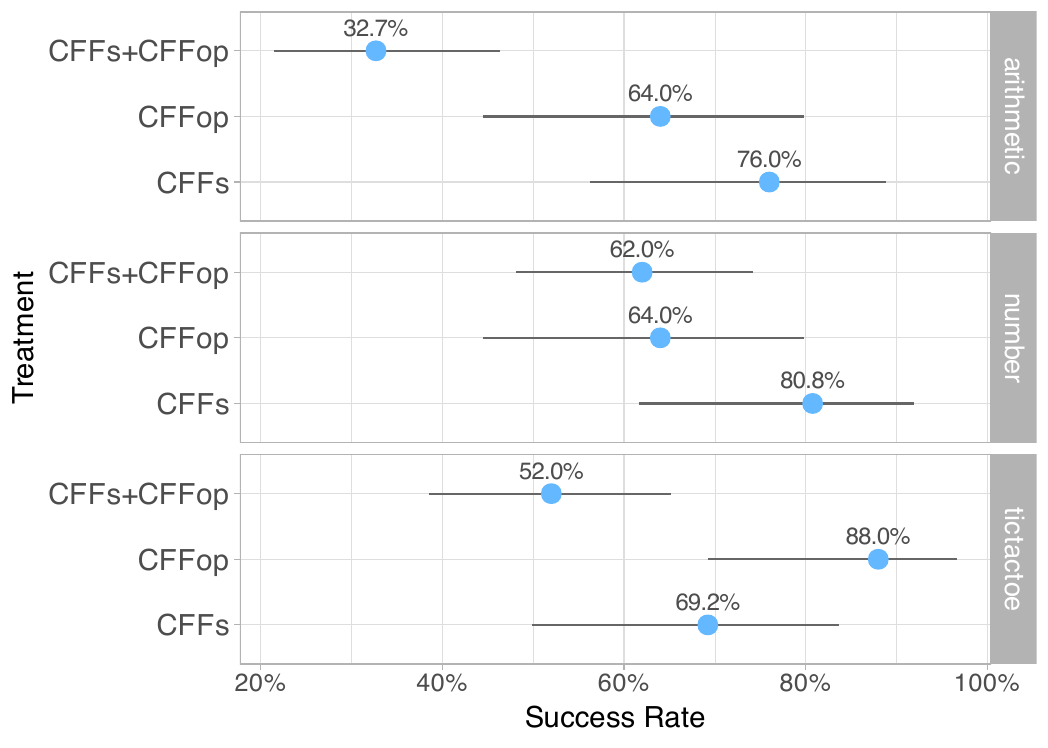}
    \caption{Success rate by obfuscation technique and application (segments indicate the 95\% CI).}
    \label{fig:success}
\end{figure}

Figure \ref{fig:success} summarises the average success rate with an indication of the 95\% confidence interval by obfuscation technique (Treatment) and \changed{analysed} application. In general, we observe how layered protection lowers the rate of success. The results of the within-subject mixed model logistic regression of success rate vs. obfuscation technique, application, period, and sequence are reported in Table~\ref{tab:lrsucc}

We draw our conclusions by applying the method reported in Section~\ref{sec:analysis_method}. We cannot reject hypothesis $H_{ss0}$, \ie no \changed{statistically} significant difference was found between the two basic obfuscation techniques.
On the other hand, we reject the hypothesis $H_{sl0}$, \ie the layered application of the two distinct obfuscation techniques has a \changed{statistically significant} impact on the success rate \wrt the basic techniques.

The odds of a successful \changed{task} when the layered techniques are applied are more than five times lower ($1/e^{-1.69}$) compared to the least effective individual technique (\VersionCFF) and almost three times lower ($1/e^{-1.04}$) than the most effective one (\VersionCFFOp). In addition, we observe a \changed{statistically} significant difference between both Number and TicTacToe applications w.r.t. Arithmetic, which has twice the odds of a successful \changed{task}. Concerning the control factors, period, and sequence, we observe no statistically significant effect; this suggests no maturation or fatigue effect was present.

\begin{table}[tb]
\caption{Mixed model logistic regression analysis for Correctness}
\label{tab:lrsucc}
\centering
\begin{tabular}{l@{}rrrr@{~}l@{}}
\toprule
Term & Estimate & Std. Error & z value & Pr(\textgreater\textbar z\textbar)&\\
\midrule
(Intercept)        &  1.134 & 0.601 & 1.887 & 0.059 & \\
\VersionCFFOp      & -0.638 & 0.558 & -1.143 & 0.253 & \\
\VersionDoubleCFF  &  1.685 & 0.431 & -3.910 & $<0.001$ & *** \\
\AppN              &  0.870 & 0.361 & 2.410 & 0.016 & *\\
\AppT              &  0.700 & 0.351 & 1.996 & 0.046 & *\\
Period             &  0.265 & 0.279 & 0.951 & 0.342 & \\
Sequence $(\VersionDoubleCFFMATH, \VersionCFFMATH)$    & -0.527 & 0.522 & -1.009 & 0.313 & \\
Sequence $(\VersionDoubleCFFMATH, \VersionCFFOpMATH)$  &  0.114 & 0.461 & 0.247 & 0.805 & \\
Sequence $(\VersionCFFOpMATH, \VersionDoubleCFFMATH  )$& -0.261 & 0.524 & -0.498 & 0.618 & \\
\bottomrule
\end{tabular}
\end{table}

\subsection{RQ2 - Time}
\label{sec:results:time}

\begin{figure}[tb]
    \centering
    \includegraphics[width=0.8\textwidth]{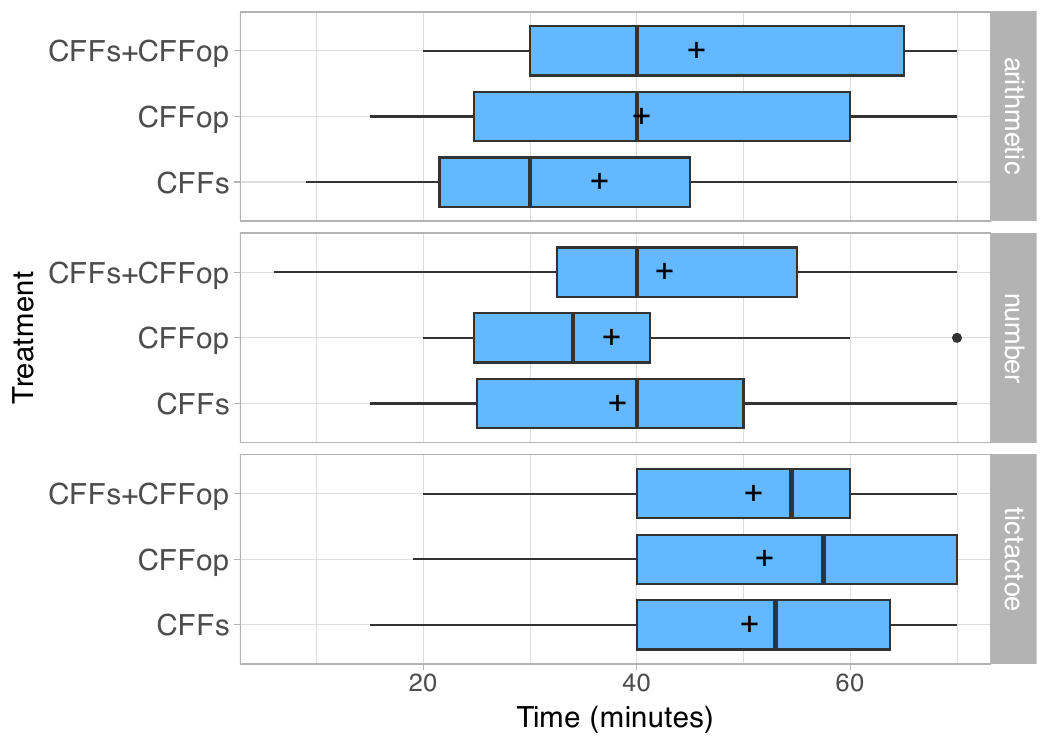}
    \caption{\changed{Code comprehension} time by Treatment and Application}
    \label{fig:time}
\end{figure}

Figure \ref{fig:time} reports a series of boxplots that summarise the distribution of time required to complete a successful \changed{task} by obfuscation technique (Treatment) and by \changed{analysed} application. Visual inspection does not allow for observing a clear trend. The mixed model results within-subject linear regression are reported in Table~\ref{tab:lrtime}.

In this case, we cannot reject hypothesis $H_{ts0}$ since there is no \changed{statistically} significant difference between the two single transformations, while we can reject hypothesis $H_{tl0}$,  \ie layering the two techniques \changed{delays in a statistically significant way} a successful task.

On average, the layering of protections increases by six minutes (~14\% of the time) to complete a successful \changed{task}. We observed a \changed{statistically} significant effect of the application TicTacToe that extended the time required to succeed by 11 minutes on average. In addition, the period had a \changed{statistically} significant negative effect, meaning that successful tasks conducted in the last period were 7 minutes shorter. This result can be seen as the consequence of a maturation effect. Students became more efficient by completing the previous tasks.

\begin{table}[tb]
\caption{Mixed model effects linear regression analysis for Time vs. Treatment and Application}
\label{tab:lrtime}
\centering
\adjustbox{max width=\textwidth}{
\begin{tabular}{@{}lrrrrr@{~}l@{}}
\toprule
Term & Estimate & Std.Error  & DF  & t-value  & p-value  &  \\
\midrule
(Intercept)        & 41.411 & 4.624 & 161.674 & 8.956 & $<$0.001 & *** \\
\VersionCFFOp      &  3.758 & 4.074 & 87.343 & 0.923 & 0.359 & \\
\VersionDoubleCFF  &  5.953 & 2.995 & 97.552 & 1.988 & 0.050 & * \\
\AppN              &  0.524 & 2.692 & 124.8 &  0.195 & 0.846 & \\
\AppT              & 11.148 & 2.720 & 124.1 &  4.099 & $<$0.001 & *** \\
Period             & -7.515 & 2.036 &  87.6 & -3.690 & $<$0.001 & *** \\
Sequence $(\VersionCFFOpMATH, \VersionDoubleCFFMATH  )$ &  0.055 & 4.524 & 169.8 &  0.012 & 0.9903 & \\
Sequence $(\VersionDoubleCFFMATH, \VersionCFFMATH)$     &  1.025 & 3.792 & 124.7 &  0.270 & 0.7874 & \\
Sequence $(\VersionDoubleCFFMATH, \VersionCFFOpMATH)$   & -2.500 & 4.545 & 168.9 & -0.550 & 0.5829 & \\
\bottomrule
\end{tabular}}
\end{table}

\subsection{RQ3 - Complexity and success rate}

\begin{figure}[tb]
    \centering
    \includegraphics[width=\textwidth]{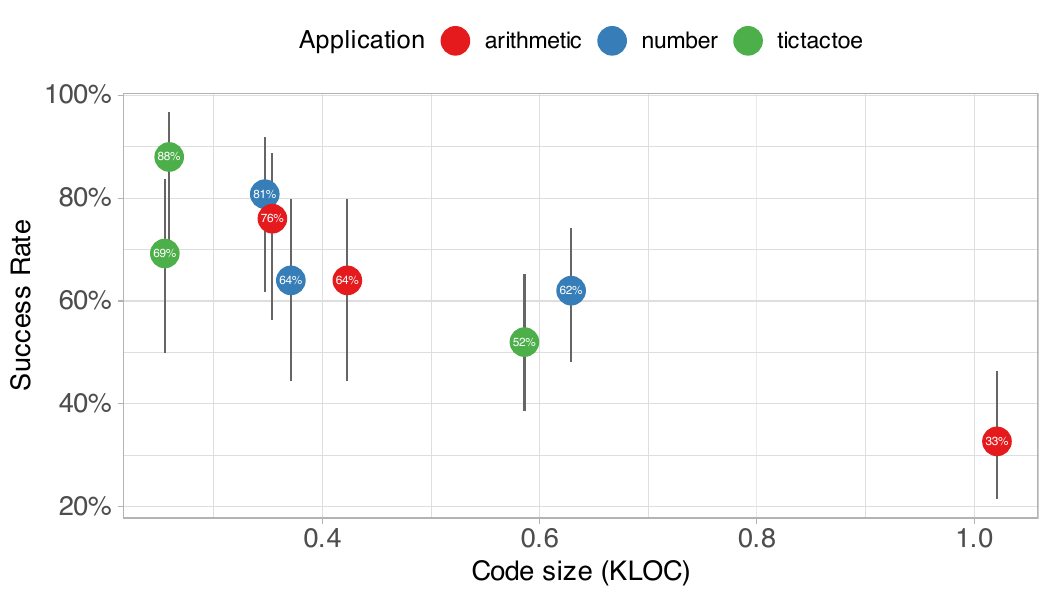}
    \caption{Success rate vs. Code size}
    \label{fig:loc_success}
\end{figure}

Figure \ref{fig:loc_success} shows the correlation between \changed{code} size, expressed as KLoC, and the corresponding average success rate with a 95\% confidence interval.
We observe a clear trend with larger applications having a lower average \changed{code comprehension} success rate. The logistic regression results of \changed{success} rate vs. \changed{code} size -- in place of \changed{treatment} and \changed{application} -- are reported in Table~\ref{tab:lrsucccode}. We observe a statistically significant effect of the \changed{code size} on \changed{code comprehension} success.
In practice, adding a thousand lines of code reduces the odds of a successful \changed{code comprehension task} by almost 24 ($e^{3.177}$) times.

\begin{table}[tb]
\caption{Mixed model effects logistic regression analysis for Success rate vs. Code Size}
\label{tab:lrsucccode}
\centering
\begin{tabular}{@{}lrrrr@{~}l@{}}
\toprule
Term & Estimate & Std.Error  & z-value  & p-value  &  \\
\midrule
(Intercept)     &  1.496 & 0.828 &  1.807 &    0.0707 & . \\
KLOC.app        & -3.177 & 0.653 & -4.865 & $<$0.0001 & *** \\
Period          &  0.244 & 0.277 &  0.880 &    0.3788 & \\
Sequence $(\VersionDoubleCFFMATH, \VersionCFFMATH)$  & 0.228 & 0.447 & 0.509 & 0.6104 & \\
Sequence $(\VersionDoubleCFFMATH, \VersionCFFOpMATH)$ & 0.390 & 0.455 & 0.857 & 0.3913 & \\
Sequence $(\VersionCFFOpMATH, \VersionDoubleCFFMATH  )$   & 0.255 & 0.445 & 0.573 & 0.5670 & \\
\bottomrule
\end{tabular}
\end{table}

\subsection{RQ4 - Complexity and time}

\begin{figure}[tb]
    \centering
    \includegraphics[width=0.9\textwidth]{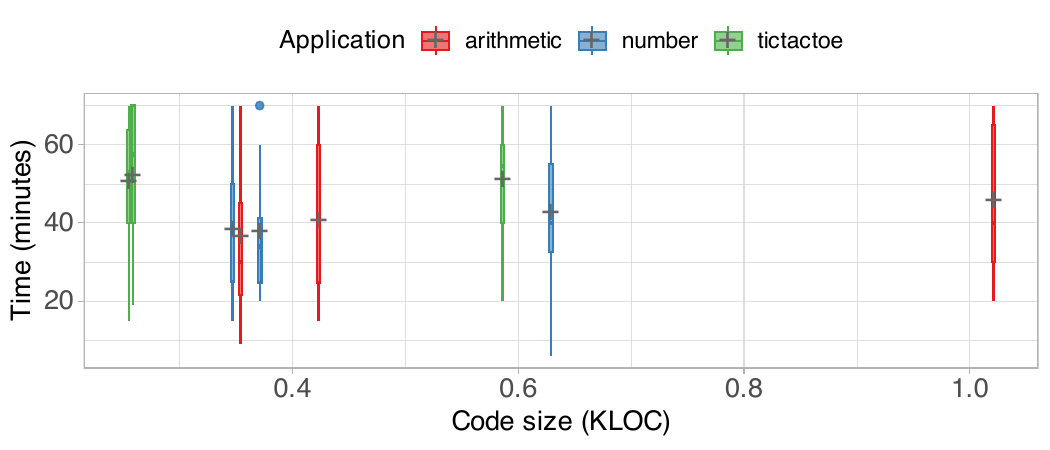}
    \caption{\changed{Code comprehension} time vs. Code size}
    \label{fig:loc_time}
\end{figure}

Figure \ref{fig:loc_time} shows the correlation between \changed{code} size, expressed as KLoC, and the time distribution of the time required to complete a \changed{task}.
Visually, we cannot see any sensible trend. The results of the linear regression of \changed{time} vs. \changed{code} size, in place of \changed{treatment} and \changed{application}, are reported in Table~\ref{tab:lrtimecode}. The test shows no \changed{statistically} significant effect of code size on the time needed to complete a \changed{task}.
Only the \changed{period} has a \changed{statistically} significant effect; as far as effect size is concerned, the tasks conducted in the second period appear shorter than those in the first one by an average of 8~minutes.

\begin{table}[tb]
\caption{Mixed model effects linear regression analysis for Time vs.\ Code Size}
\label{tab:lrtimecode}
\centering
\adjustbox{max width=\textwidth}{
\begin{tabular}{@{}lrrrrr@{~}l@{}}
\toprule
Term & Estimate & Std.Error  & DF  & t-value  & p-value  &  \\
\midrule
(Intercept)     & 47.605 & 3.746 & 185.4 & 12.706 & $<$0.0001 & *** \\
KLOC.app        &  1.530 & 5.330 & 117.5 &  0.287 &    0.7745 & \\
Period          & -8.160 & 2.331 &  97.6 & -3.501 &    0.0007 & *** \\
Sequence $(\VersionDoubleCFFMATH, \VersionCFFMATH)$ & -0.963 & 3.818 & 124.9 & -0.252 & 0.8013 & \\
Sequence $(\VersionDoubleCFFMATH, \VersionCFFOpMATH)$   &  0.358 & 3.838 & 109.2 &  0.093 & 0.9259 & \\
Sequence $(\VersionCFFOpMATH, \VersionDoubleCFFMATH  )$   &  2.090 & 3.805 & 110.8 &  0.549 & 0.5840 & \\
\bottomrule
\end{tabular}}
\end{table}

\subsection{Cofactors: skills}

The main co-factor that we considered when balancing the composition of the groups in the experiment was the C language skill.
As a confirmatory check, we thus analysed the relationship between the measures of C Skill -- C Skill home and C Skill live described in section \ref{sec:variables} -- and the two output variables, success rate and time to complete \changed{a code comprehension task}.

\begin{figure}[tb]
    \centering
    \includegraphics[width=0.49\textwidth]{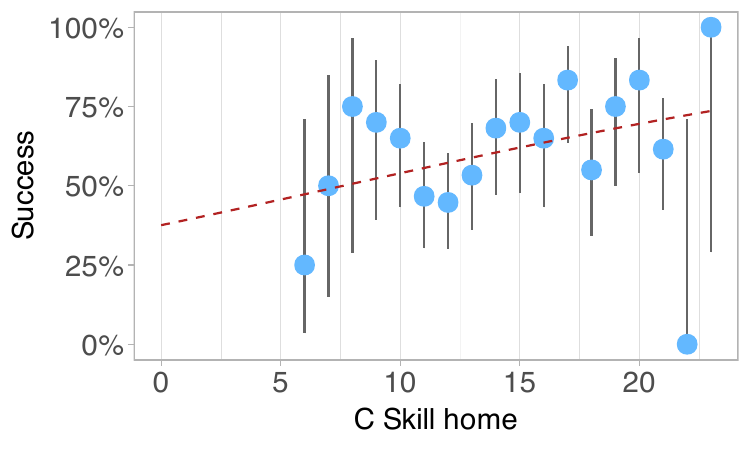}
    \includegraphics[width=0.49\textwidth]{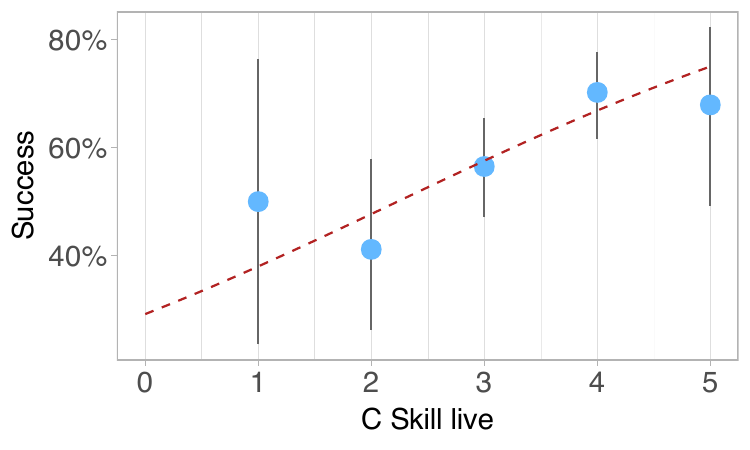}
    \caption{Success rate vs C Skill.}
    \label{fig:skill_success}
\end{figure}

Figure \ref{fig:skill_success} reports the success rate vs. the two skill measures, together with a fitted logistic regression curve. We can visually observe a trend that is confirmed by the p-values of the ANOVA: 
C Skill Home $p = 0.0319$, C Skill Live $p = 0.00264$.

\begin{figure}[tb]
    \centering
    \includegraphics[width=0.49\textwidth]{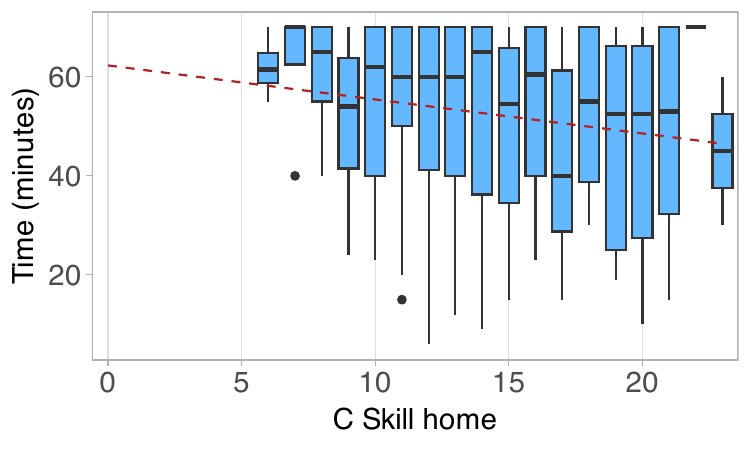}
    \includegraphics[width=0.49\textwidth]{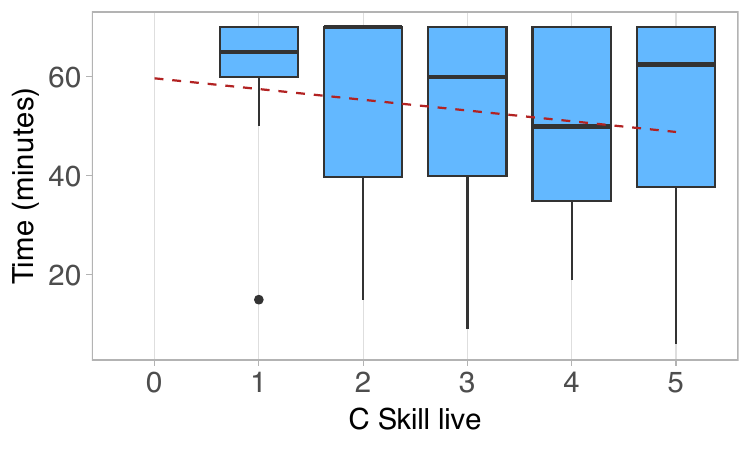}
    \caption{Success rate vs C Skill.}
    \label{fig:skill_time}
\end{figure}

Figure \ref{fig:skill_time} reports the time to complete the \changed{task} vs.\ the two skill measures, together with a fitted regression line. We can visually observe a weak trend. The ANOVA's corresponding p-values are p = 0.0107 for C Skill Home and p = 0.056 for C Skill Live.

In summary, we can confirm that C skill affects the success rate and partially the time, though in both cases, it cannot explain much; it is just an additional co-factor, which is, in any case, taken into account by the within-subjects models described above.

\subsection{Models comparison}

The RQs led us to define two distinct models explaining both the \changed{code comprehension} success rate and time to complete. \changeda{We compare them here, also reporting two additional models for each output built using the two C Skill measures.}

\begin{figure}[tb]
    \centering
    \includegraphics[width=0.85\textwidth]{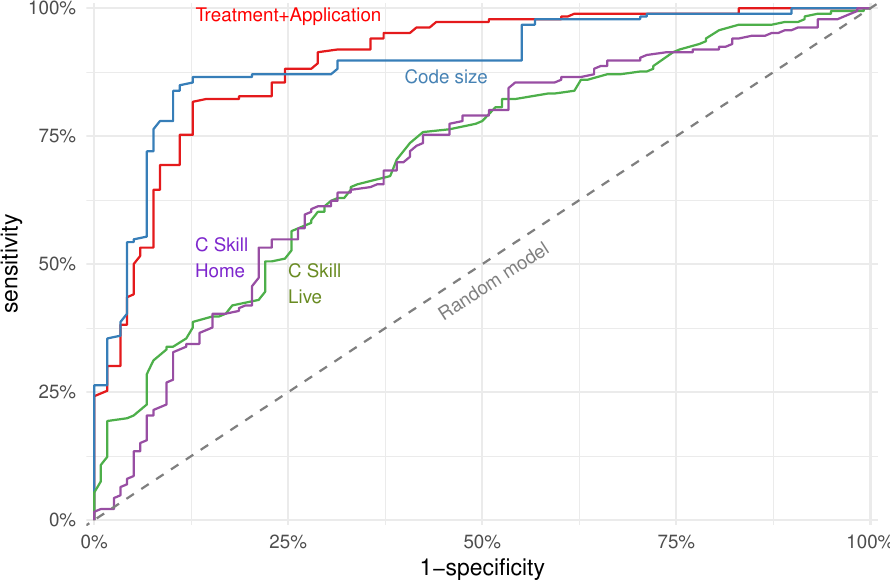}
    \caption{ROC curves for the Success models}
    \label{fig:ROCs}
\end{figure}

\textbf{Success rate models.} The ROC curves for the alternative logistic regression models -- \ie, the ones in Tables~\ref{tab:lrsucc} and~\ref{tab:lrsucccode} and similar models for the C Skills -- for success rate are reported in Figure~\ref{fig:ROCs}. The AUC for the two upper curves are $0.897$ for the model that considers treatment and application, and $0.894$ for the model with code size.
DeLong's test for two correlated ROC curves yields a p-value$=0.7224$. Therefore, we cannot reject hypothesis $H_{ms0}$. The performance difference between the two models is quite small and not statistically significant.
We also observe that for small values of false positive rate -- \ie 1-specificity -- the model based on code size outperforms the one based on \changed{treatment and application}. The AUC for the C Skill home model is 0.702, whereas the AUC for the C Skill live model is 0.711. We can clearly observe from the diagram that these latter two models are statistically different from the former ones.

\textbf{Time models.} The accuracy of the alternative models for time to complete a \changed{code comprehension task} can be compared using two information criteria, AIC and BIC, \changeda{reported in Table~\ref{tab:lrtIC}}. The lower the value of the indexes, the better the accuracy.
The models' accuracy is relatively similar, with a slight edge for the model based on \changed{treatment} and \changed{application}.

\begin{table}[tb]
\caption{Accuracy information criteria values}
\label{tab:lrtIC}
\centering
\begin{tabular}{@{}lrr@{}}
\toprule
Model                 & AIC       & BIC \\
\midrule
Treatment+Application &  1576.156 &  1611.639 \\
Code size             &  1594.495 &  1620.301 \\
C Skill home          &  2605.213 &  2646.101 \\
C Skill live          &  2607.738 &  2648.626 \\
\bottomrule
\end{tabular}
\end{table}

\subsection{Participant perception}
\label{sec:postq}

\begin{figure}[tb]
    \centering
    \includegraphics[width=\textwidth]{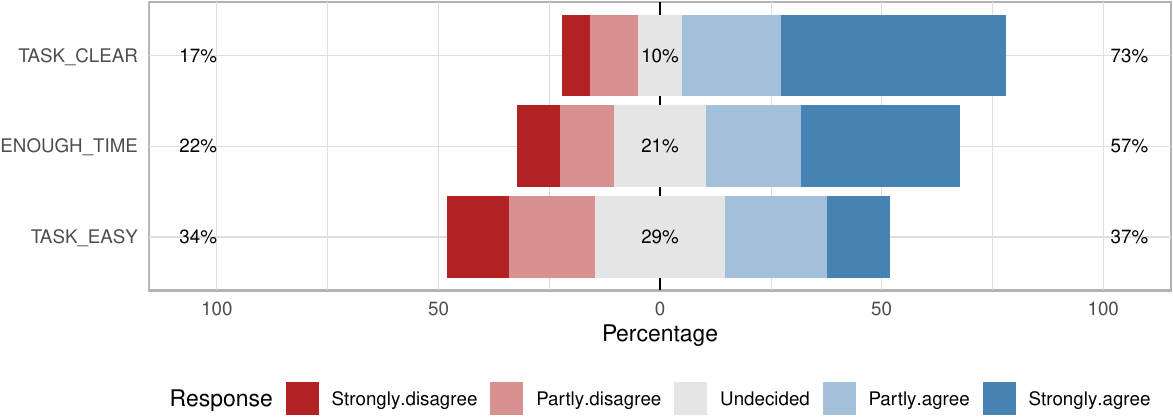}
    \caption{Participant responses to post-experiment questionnaire}
    \label{fig:postq}
\end{figure}

Figure \ref{fig:postq} reports the distribution of participants' answers to three key questions, \changeda{concerning} the clarity of the task they had to perform ($\mathrm{TASK\_CLEAR}$), the availability of enough time ($\mathrm{ENOUGH\_TIME}$), and the task being easy to perform ($\mathrm{TASK\_EASY}$). We observe that the task was clear for a large majority of participants, and they considered the time sufficient in 57\% of the tasks. A less clear picture concerns the ease of performing the task, which is consistent with the overall success rate of tasks (61\%).

\section{Discussion}
\label{sec:discussion}

Our analysis has identified several points worth discussing. Either they shed light on the \changed{code comprehension} process and \changed{obfuscation} potency or present interesting aspects that deserve further research, studies, and experiments.

\textbf{Correlation with the objective metrics.}
This experiment has provided the first scientific evidence of a correlation between objective metrics, typically used to evaluate code quality, and the comprehension task. 
By digging deeper into the reasons for the application's influence on both success rates and time, we noted that the success rates and time strongly depend on the code complexity of the applications being tampered with.
This correlation was predicted by Collberg \etal~\cite{collberg1997taxonomy}, but no evidence was available to support it.
Indeed, the analysis has shown that objective metrics are as effective in predicting the success of a task as application and treatment combined. We cannot isolate the metric that has the most important impact, as the treatments on the considered applications increased the values of the metrics, that is, for all the objects
$$\mu(\mathrm{vanilla})<\mu(\VersionCFFOpMATH)<\mu(\VersionCFFMATH)<\mu(\VersionDoubleCFFMATH)$$

\textbf{Dig into the role of the metrics and their relations with the human brain.}
More experiments would be needed to estimate the individual impacts of metrics. Experiments with more heterogeneous techniques can help determine whether some protections also compel attackers to modify their attack strategies. In particular, it would be interesting to evaluate when attackers abandon static techniques (\ie reverse engineering tools like Radare2) to use dynamic techniques (\eg debuggers). Moreover, it would be key to determine the characteristics that make code less understandable by correlating them precisely to physiological aspects of the human brain.

\textbf{Layered protection.}
\changeda{This experiment provides evidence to support the current practice of applying multiple protections to the same pieces of code. Layered protection reduces the success rate in understanding code in a given time frame by 3-6 times (see Table~\ref{tab:lrsucc}).
Moreover, it slightly delays (5\%-14\%) the subjects who succeed.
We cannot determine whether layered protection is effective because of the differences introduced in the protected code by using diverse techniques, or whether it simply increases the complexity metrics more than a single protection. 
Our intuition indicates that different transformations may impact different aspects of comprehension. Hence, more experiments would be needed to prove this claim, using protection techniques less homogeneous than the ones we used in our experiment (\eg anti-tampering, data protection, renewability techniques, local vs. remote approaches). Using diverse techniques could also highlight differences in the attack strategies involved by the presence of protections.}

\textbf{App logic may have an impact.}
The experiments indicated that a specific application, tictactoe, was more complex to tamper with than the increase in complexity could explain.
Further experiments are necessary to verify this hypothesis and identify which elements of the application's logic made the task more challenging for the subjects. It would also be insightful to correlate these elements with human cognitive abilities. Moreover, it would be interesting to define experiments that can isolate the task of locating the areas to modify from the complexity of the changes to be made, to consider the task as successful. Indeed, our replication package could serve as a starting point for further investigations into the role of business logic in reverse engineering.

\textbf{Ensuring enough successes.}
\changeda{One crucial aspect to consider when designing controlled experiments is the number of subjects involved and their actual chances of success in the allotted time.} The risk is that hypotheses cannot be confirmed or discarded with statistical significance.
The fact that only a limited number of subjects succeeded in their tasks limited our ability to precisely correlate time and treatment.
Indeed, treatments were chosen among the obfuscation techniques implemented by Tigress, considering the subjects' expertise. The purpose was to allow a reasonable number of successes in the time allotted for the experiment. We manually inspected applications protected by several techniques and discarded those we estimated were too complex. 

\textbf{Subject abilities matter.}
The results in the C programming have a loose correlation with success and time. A trend is visible, but the skill variables are not statistically significant.
Involving professional hackers may be necessary if \changed{task complexity} increases, also considering the need to ensure a proper number of successes.

\textbf{Subject experience matters.}
We noticed that in the second period, the subjects are faster (about 7.5 minutes), which can be attributed to a learning/maturation effect. \changeda{Additionally, some students reported comments like ``the code looked similar to the first program.'', suggesting that subjects were able to learn something about obfuscation techniques in just a 4-hour session. However, we cannot scientifically determine the extent of this learning based on the available data. Professional pentesters could perform considerably more quickly, which is a concern, as an application is considered compromised as soon as the first attacker cracks it.} Correlating the speed of expert pentesters with the results from our empirical experiments involving students may be essential for designing effective protections and assessing the resilience of systems against attacks.

\textbf{Selecting the right subjects for assessing software protection.}
Finally, we report that the analysis of the processes, as reported in \ref{sec:process_mining}, indicated that some subjects who failed were probably unable to approach the assigned task, despite the skill tests indicating that all had a sufficient background.
This consideration raises a discussion on methods to select more suitable subjects for experiments to assess the effectiveness of software protection and other cybersecurity-related tasks.

%%%%%%%%%%%%%%%%%%%%%%%%%%%%%%%%%%%%%%%%%%%%%%%%%%%%%%
\section{Conclusions and future work}
\label{sec:conclusions}
In this work, we have reported the data collected in a controlled experiment on the effectiveness of layering two widely adopted code obfuscation techniques: \cff and \ops. \changeda{The experiment involved 152 MSc students as subjects, who performed three different tasks on three different applications}. In particular, after executing a preliminary task on a vanilla application, the subjects performed tasks on applications obfuscated with two different versions of \cff and another on an application obfuscated by layering both versions. \changed{All tasks required the subjects to understand the obfuscated code, demonstrating their success by fixing a trivial bug in the targeted application.}

\changeda{By analysing the results, we show that layering protection is highly effective; it significantly correlated with the task's success, reducing the odds of a successful \changed{code comprehension} by up to 5.4 times with respect to the same application protected with only one of the obfuscation techniques. However, it showed no statistically significant correlation with the code comprehension time. We also noted differences in the \changed{code comprehension} success and time depending on the application to \changed{analyse}.}
Moreover, we highlighted a learning effect, as the time to succeed in the second main task was significantly shorter than the time to succeed in the first.

\changeda{To the best of our knowledge, this is the first empirical experiment able to correlate code complexity metrics and the \changed{code comprehension}'s success rate, which constitutes the basis of the software protection potency metric introduced by Collberg\etal~\cite{collberg1997taxonomy}.} Interestingly, the complexity of an application only filters out the people who succeed, while having a minor impact on the time it takes to successfully complete the task: fewer subjects succeeded, but required approximately the same amount of time.

Following this work, we plan to organise more controlled experiments. \changeda{First, we assess the impact of layering additional protections, for example, the typical layering of obfuscation over anti-tampering techniques (\eg remote attestation~\cite{viticchie2016reactive} or code guards~\cite{codeGuards}), obfuscating the code that verifies and reacts to unauthorised modifications of the protected application.} We also plan to involve professional hackers to assess if layering maintains its impact on code comprehension odds when advanced attack tools and techniques are used (\eg concolic analysis, taint analysis, professional debuggers). \changeda{Involving expert subjects would help assess the impact of using source or binary code as experimental objects, clarifying the effects of employing students as experimental subjects in empirical assessment studies of software protection.}

\changeda{Furthermore, future studies may leverage larger datasets, comprising programs with increasing complexity and size, to construct regression or machine learning models that estimate expected code comprehension time or success probability as a function of program size, obfuscation configuration, or code structure. Such models would allow developers to choose protection strategies commensurate with the assets' risk and value.}

Finally, future works should investigate how the increased difficulty induced by code obfuscation relates to subsequent tampering activities performed on the same code, examining whether specific obfuscation strategies (\eg \cff, \ops) differentially increase the effort required for tampering, and whether this effect persists across attacker profiles, task types, and time constraints. Establishing these relationships would help clarify how obfuscation contributes to tamper-resistance towards cost-effective software protection.

\section*{Data Availability Statement}
Data supporting this study are openly available from GitHub at \url{https://github.com/daniele-canavese/empirical-obfuscations}.

\section*{Acknowledgments}
%For editorial team: it is VERY important that the ack to the SERICS project is published exactly as reported
This work was partially supported by project SERICS (PE00000014) under the NRRP MUR program funded by the EU - NGEU. This work was partially supported by ICO, Institut Cybersécurité Occitanie, funded by Région Occitanie, France, and by the European research projects H2020 LeADS (GA 956562), Horizon Europe DUCA (GA 101086308), ARN TrustInClouds, and CNRS IRN EU-CHECK.

\section*{Compliance with Ethical Standards}
\paragraph{Human participants and educational context} The study involved students enrolled at the Politecnico di Torino in Italy. The activity formed an optional component of a course assessment. Participation was voluntary, and students could opt out at any time without penalty.
\paragraph*{Informed participation and confidentiality} Students were informed of the aims and procedures in advance. Collected data were restricted to task results and minimal task-related metadata necessary for teaching administration and evaluation. Research analyses used de-identified and aggregate data; no personal identifiers are stored, reported, or published.
\paragraph*{Data protection} Data processing complied with the Art. 13 of EU Regulation 2016/679 (General Data Protection Regulation) and the Politecnico di Torino Students’ Data Privacy Policy. The University acts as Data Controller; Data Protection Officer contact details are available on the institutional privacy pages.
\paragraph*{Conflict of interest} The authors declare no conflicts of interest.

\bibliographystyle{elsarticle-num}
\bibliography{biblio}

%%%%%%%%%%%%%%%%%%%%%%%%%%%%%%%%%%%%%%%%%%%%%%%%%%%%%%
\parpic{\includegraphics[width=0.9in,clip,keepaspectratio]{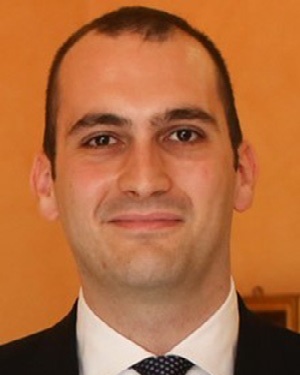}}
\noindent\textbf{Leonardo Regano} received an M.Sc. degree in 2015 and a Ph.D. in Computer Engineering in 2019 from Politecnico di Torino, where he worked as a research assistant for eight years. He is currently an assistant professor at the Department of Electrical and Electronic Engineering, University of Cagliari (Italy). His current research interests focus on software security, artificial intelligence and machine learning applications to cybersecurity, security policies analysis, and software protection techniques assessment. 
\vspace{2em}

\parpic{\includegraphics[width=0.9in,clip,keepaspectratio]{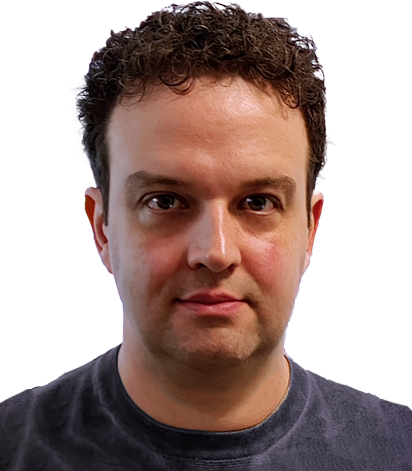}}
\noindent\textbf{Daniele Canavese} received an M.Sc. degree in 2010 and a Ph.D. in Computer Engineering in 2016 from Politecnico di Torino, where he worked as a research assistant for more than ten years. He is currently a post-doc researcher at the IMATI (Istituto di Matematica Applicata e Tecnologie Informatiche), in Genova (Italy). His current research interests include using artificial intelligence and machine learning techniques for security management, software protection systems, public-key cryptography, and models for network and traffic analysis.
\vspace{1em}

\parpic{\includegraphics[width=0.9in,clip,keepaspectratio,trim={0 15mm 0 0}]{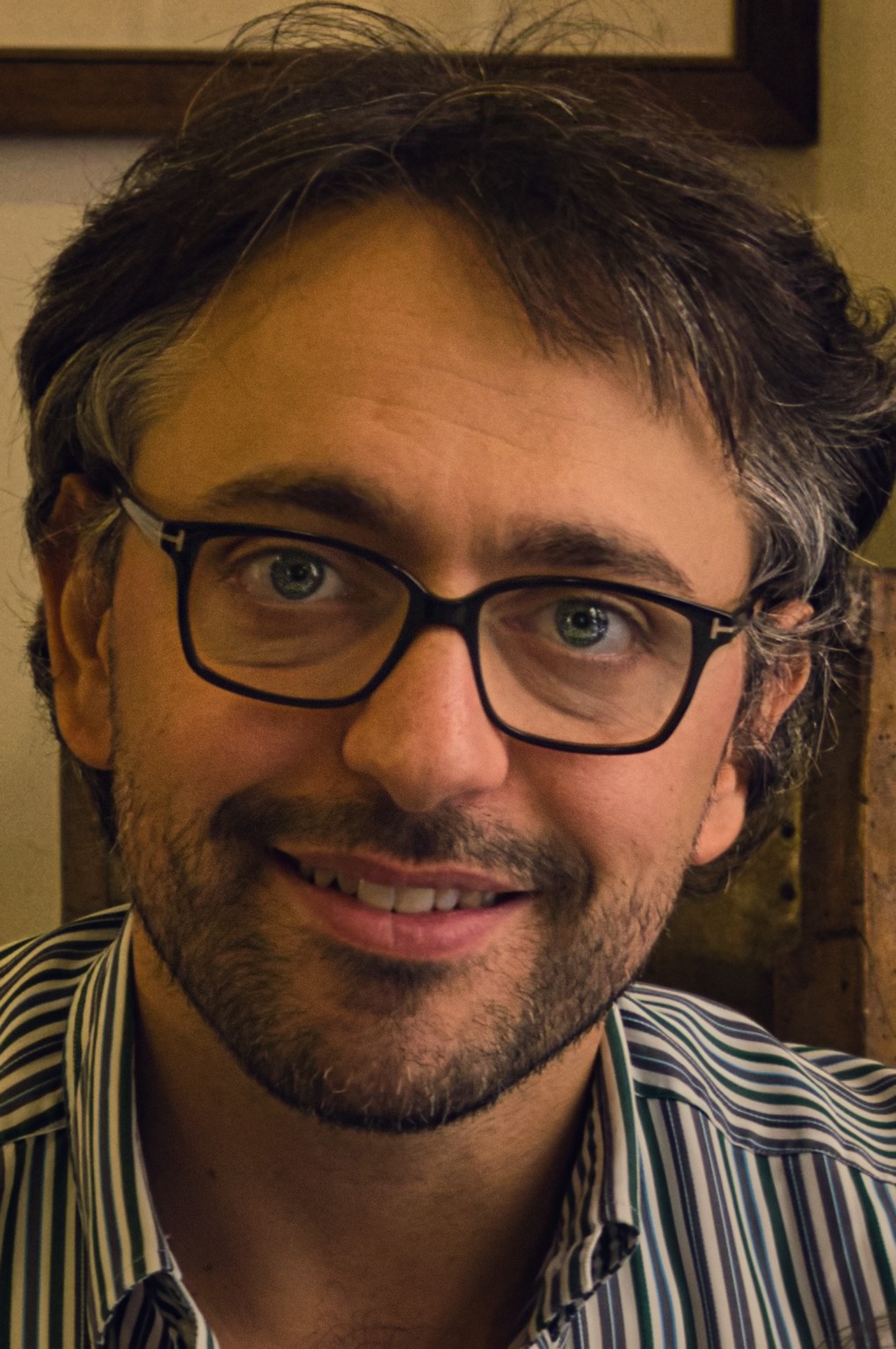}}
\noindent\textbf{Cataldo Basile} is an associate professor at the Politecnico di Torino, from which he received an M.Sc.in 2001 and a Ph.D. in Computer Engineering in 2005. His research concerns software protection, software attestation, policy-based security management, and general models for detecting, resolving, and reconciling security policy conflicts.
\vspace{4em}

\parpic{\includegraphics[width=0.9in,clip,keepaspectratio]{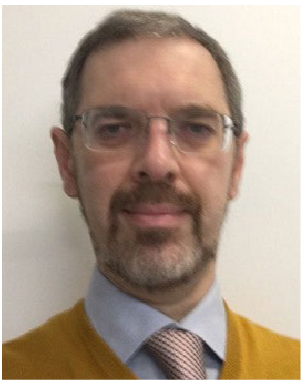}}
\noindent\textbf{Marco Torchiano} received the M.Sc. and Ph.D. degrees in computer engineering from Politecnico di Torino. He is a full professor in the Department of Control and Computer Engineering at Politecnico di Torino, Italy. His current research interests include green software, UI testing methods, open-data quality, and software modelling notations. The methodological approach he adopts is that of empirical software engineering.

%He has been a Postdoctoral Research Fellow with the Norwegian University of Science and Technology (NTNU), Norway. He is a member of the software engineering committee of UNINFO (part of ISO/IEC JTC 1). He is the Author or a Co-Author of over 140 research papers published in international journals and conferences, a Co-Author of the book titled Software Development—Case Studies in Java (Addison-Wesley), and a Co-Editor of the book titled Developing Services for the Wireless Internet (Springer). Recently, he was a Visiting Professor at Polytechnique Montréal studying software energy consumption.

\appendix
% \input{appendix}
% %%%%%%%%%%%%%%%%%%%%%%%%%%%%%%%%%%%%%%%%%%%%%%%%%%%%%%
\section{Process mining}
\label{sec:process_mining}
We collected 456 reports describing the attack procedure followed by subjects on their attack tasks. They form the basis used to perform process analysis.
This analysis aimed to extract information about the attack strategies and the approaches subjects used when performing their attack tasks.
The high-level objective is obtaining confirmations of the results in Section~\ref{sec:results} and more insights into how subjects performed their tasks.

More in detail, we wanted to determine if subjects followed general attack strategies when performing attack tasks and if subjects who did not succeed were following a different approach from the subjects who instead succeeded.
Moreover, we wanted to understand whether the attack process changed in the presence of the protections (\ie vanilla applications used during the warm-up vs. protected applications in period~1 and~2) and if these changes were affecting the time required to complete the attack. 
We also wanted to see if the treatment impacted the process, understand why layered protections were more effective, and the role of complexity metrics.

Finally, we wanted to investigate if the attack process was different depending on the target applications and understand why attacking TicTacToe was more difficult.

\subsection{Preparing the reports with a closed coding approach}

The reports consisted of a sequence -- up to 20 -- attack steps, each composed of three fields:
\begin{itemize}
    \item \textit{assets}, a free text form describing the assets involved in the attack step;
    \item \textit{description}, an open description of the activities performed, their considerations, what they observed, etc.
    \item \textit{label}, the categorization of the step based on the fixed list of attack steps in the ontology the subjects received before the experiment.
\end{itemize}
The sequences of attack steps have been named in the literature as \textit{attack path} \cite{reganoMeta}.
The reports have been processed using a closed coding approach using the concepts from the ontology provided to the students. This ontology included 32 attack steps from a bigger taxonomy, including 169 concepts, developed to describe hackers while performing attack tasks \cite{emse2018}.

Initially, a collegiate evaluation of the coherence of the data available in the reports was performed.
A preliminary analysis allowed evaluating the information in the \textit{assets} field as not useful or of very little usability for our analysis.
The reports contained vague indications (just the name of the whole file they modified: useless since only one file was to modify) or erroneous (incorrect names of functions and variables). Therefore, this field was not considered in the rest of the process analysis.

Then, we started processing the \textit{label} assignments and noted incoherencies with the \textit{description} field. A non-negligible number of \textit{labels} were not consistent with the activity described in the free-text description field.
We attributed this mainly to a problem of limited clarity of the material provided to the participants, emphasized by the limited subjects' background in software protection; moreover, we cannot exclude that they did not properly study the material provided.

For this reason, we decided to perform an additional step to correct label assignments that were clearly wrong. 
Every attack step report was independently processed by at least two reviewers, from the authors of this paper\footnote{Stefano Alberto also reviewed reports, but he left our institution before starting to write this paper.}
Every reviewer could only perform the following changes to the reports:
\begin{itemize}
    \item Replace the attack step label with another one he was considering more appropriate.
    For instance, the following comment, ``\textit{I fixed the code changing the shift from 47 to 48,}'' was initially annotated as ``defeat protection''. After this review, it was assigned to ``tamper with code statically''.
    
    \item Add an attack step in the attack path when the text implies it. For instance, ``\textit{Simply found a strange thing at line 281, I couldn't understand why there was an \& 4039822362U [..]. So I removed it, and the program started working.}'' was initially annotated as ``tamper with data'', but an ``analyse attack results'' step was added, as the subject also assessed the results of the modifications.
    
    \item Delete redundant attack steps only when subjects were reporting two consecutive attack steps with the same label to perform the same task on the same asset (instead, the same task on different assets or parts of the applications was considered acceptable). For instance, we found cases of multiple ``identify assets by naming scheme'' to mark consecutive searches based on file and asset names or ``tamper with data'' when several variables were changed.

    \item Mark the entire attack path as incoherent when reports contained severe inconsistencies (\eg a non-reconcilable sequence of incoherent steps without proper free text), as we refrained from adding any personal judgements or corrections that were not directly expressed in the free text field.

\end{itemize} 

Then, after the independent reviews, the involved reviewers made a bilateral call to find an agreement on the proposed changes and generate the final reports.
When both reviewers marked a report as incoherent, it was removed from the rest of the process analysis. In the end, 83 (out of 456) reports were excluded from the next steps of the process analysis: 29/152 of Period~1, 25/152 of Period~2, and 29/152 of Period~3.
We observed a similar number of exclusions during the three periods; this means that students did not improve their writing reports during the experiment. They probably needed more experience to be able to fill in the report properly. We included the attack paths in the replication package.

%%%%%%%%%%%%%%%%%%%%%%%%%%%%%%%%%%%%%%%
\subsection{Process mining on the attack paths}

The reports have been analyzed with an open-source process mining library (pm4py\footnote{\url{https://processintelligence.solutions/pm4py}}) that implements a Heuristic Mining algorithm \cite{Weijters2006ProcessMW}.
Heuristics Miner is an algorithm that acts on the Directly-Follows Graph, providing a way to handle noise and find common constructs.
This algorithm finds a process model that describes the order of events/activities that happen during the execution of a process and has been applied to the attack paths.
In practice, given a set of reports, it outputs a graph describing the process.

The process graphs resulting from process mining are formed by: 
\begin{itemize}
    \item  a set of nodes $a\in \mathcal{A}$, where each $a$ is an attack step, associated with a function $\alpha(a)$ to the number of times that $a$ appeared in the report;
    \item a set of edges $e=(a_1,a_2)\in \mathcal{E}=\mathcal{A}\times\mathcal{A}$, associated with a function $\epsilon(e)$ to the number of times that the sequence $a1,a_2$ appeared in the report.
\end{itemize}

We have considered three variables to categorize the reports and generate specific process graphs (see Section~\ref{sec:variables}):
\begin{itemize}
    \item \textit{object}, \ie the three application to tamper with;
    \item \textit{treatment}, \ie the three protections used and the vanilla applications that have been provided to subjects during the warm-up experiment;
    \item the \textit{succeeded} variable (\eg succeeded, failed, both).
\end{itemize}

The reports were divided into $3\times3$ (objects$\times$succeded) + $4\times3$ (objects$\times$succeded) + $3$ (succeeded) = $24$ sets, corresponding to all the combinations of objects, treatments, and success information and a set including all the reports.

\begin{figure}
    \centering
    \includegraphics[width=\textwidth]{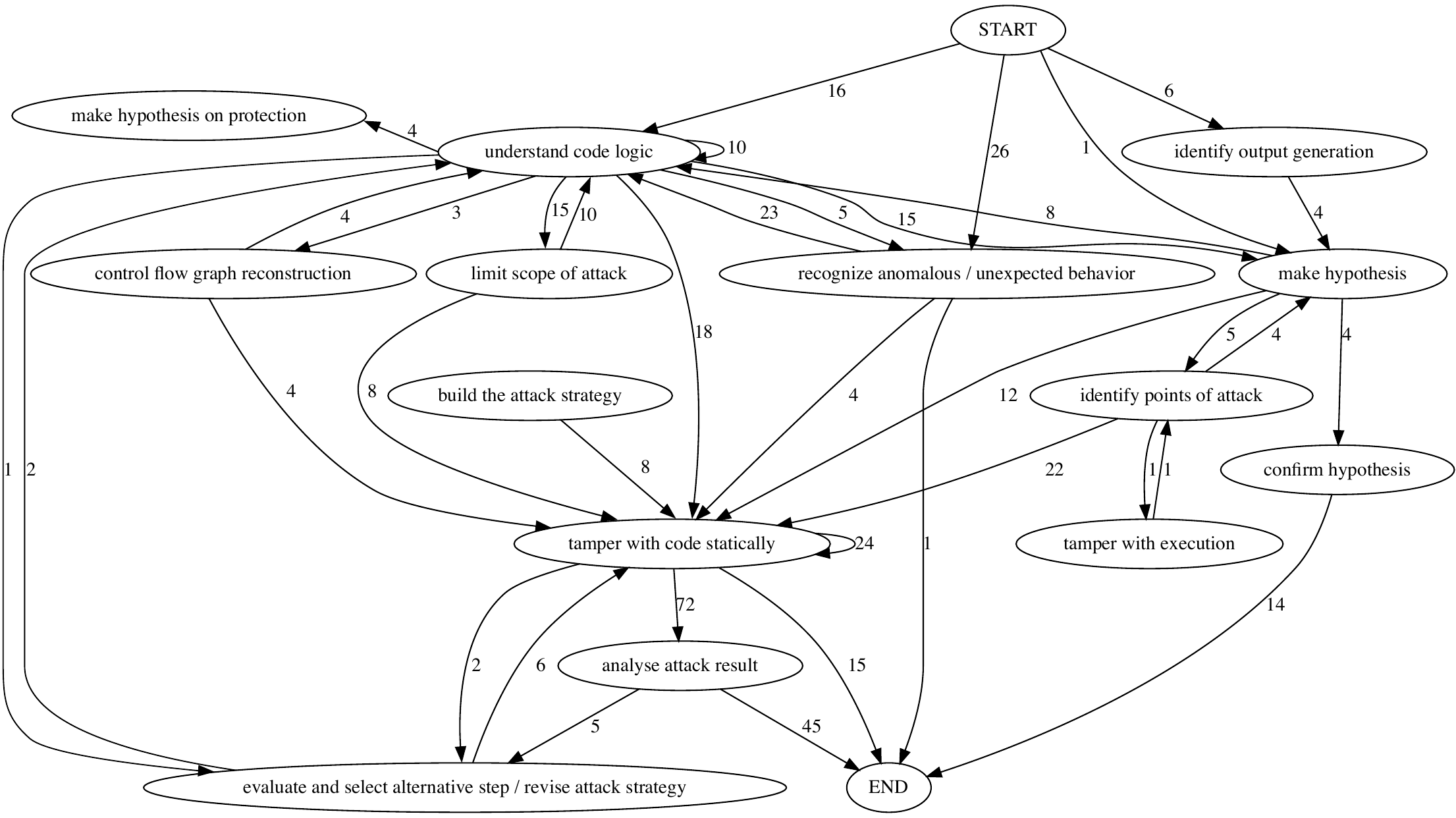}
    \caption{Process for succeeded tasks on the arithmetic app ($(\delta=0.8,\alpha=0,\epsilon=0)$).}
    \label{fig:proc_succ_80}
\end{figure}

The pm4py library implementation of the Heuristic Miner algorithm accepts different parameters to fine-tune the process mining by filtering nodes and edges based on noise thresholds.
For this analysis, we have used the following parameters:
\begin{itemize}
    \item DEPENDENCY\_THRESH ($\delta\in [0,1]$), the threshold for the strength of the directly follows dependency, which is used to filter the less evident relations. The higher the number, the more nodes and edges are filtered.
    \item MIN\_ACT\_COUNT ($\alpha$), the minimum number of occurrences of an activity to be considered, used to prune the attack steps that appeared too few times.
    \item MIN\_DFG\_OCCURRENCES ($\epsilon$) the minimum number of occurrences of an edge, served to prune the sequences of two consecutive attack steps that appear too few times.
\end{itemize}
After experimenting with the values, we have decided to use the following combinations 
\begin{itemize}
    \item $(\delta=0.5,\alpha=2,\epsilon=2)$ and $(\delta=0.5,\alpha=3,\epsilon=3)$ for the graphs used by automated procedures, as they are relatively large; for example, the graph for the arithmetic application contains 24 nodes and 81 edges.
    \item $(\delta=0.5,\alpha=0,\epsilon=0)$, $(\delta=0.65,\alpha=0,\epsilon=0)$, and $(\delta=0.8,\alpha=0,\epsilon=0)$ for manual inspection, as the generated graphs were simple enough to be analysed by humans (see Figures~\ref{fig:proc_succ_80}).
\end{itemize}

Overall, our analysis included $5\times 24 = 120$ process graphs, also available in the replication package.

Analyzing the process graphs obtained with the process mining revealed several interesting facts.

It is impossible to infer the existence of a single attack strategy for performing the attack tasks.
However, two main attack approaches emerged: the \textit{backward} one, where the subject started from the output and traced back in the source code to the point where the bug originated and 
the \textit{forward} one, where the subject started from the \lstinline|main()| function and followed the execution flow (with a debugger) until the bug was found.
We manually annotated the attack paths with forward/backward labels; there is no significant evidence to suggest that one approach is better than the other for successfully solving the tasks.

Moreover, we have qualitatively noted that graphs of successful attacks are a little more complex (a few more nodes and edges, a more structured sequence of steps vs.\ a flatter graph) than the graphs of failed attacks. Moreover, it was evident that the processes followed by those who succeeded looked very similar.
Hence, we have introduced a metric (a distance) to measure the similarity between two graphs and confirm our feeling.
The metric introduced to measure the graph similarity is based on the Graph Edit Distance (GED) \cite{GED}.
GED is defined as the number of edge/node changes needed to make two graphs isomorphic.
The computation of the exact value of the GED is an NP-hard problem.
For this reason, we used the optimize\_graph\_edit\_distance available in the \textit{networkx} Python library\footnote{\url{https://networkx.org/}} to compute an approximated value of the distance.

\begin{figure}[tb]
    \centering
    \includegraphics[width=\textwidth]{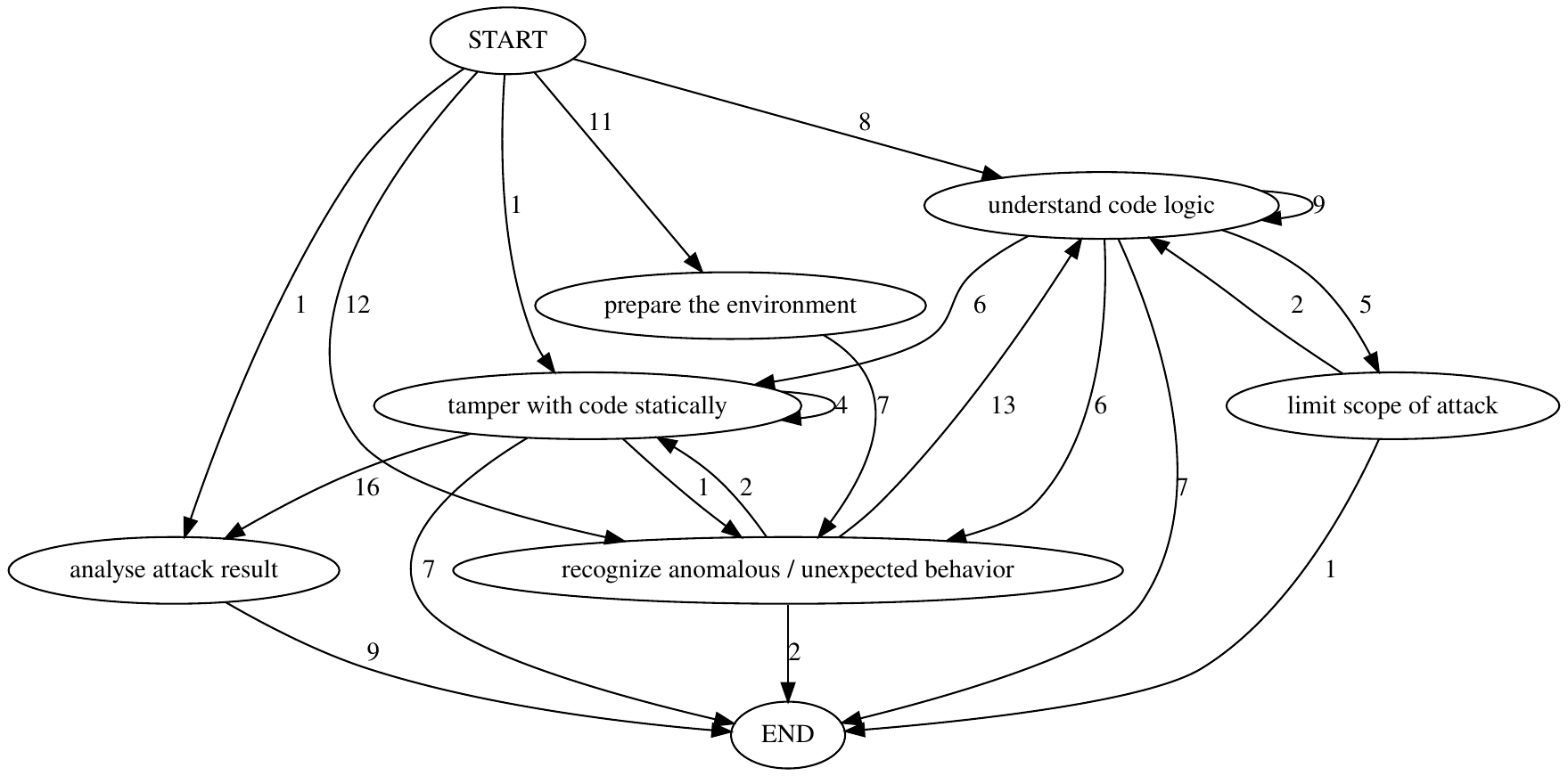}
    \caption{Process for failed tasks against the arithmetic app $(\delta=0.8,\alpha=0,\epsilon=0)$.}
    \label{fig:proc_fail_80}
\end{figure}

\begin{table}[tb]
    \centering
    \begin{tabularx}{.8\textwidth}{lcX}
        \toprule
        Action & Default Cost & Modified Cost\\
        \midrule
        Node $n$ insertion/deletion & 1 & $\alpha(n)/|\mathcal{A}|$ \\
        Edge e deletion/insertion & 1 & $\epsilon(n)/|\mathcal{E}|$ \\
        Node substitution $n_1\to n_2$ & 0 & $|\alpha(n_2)/|\mathcal{A}|-\alpha(n_1)/|\mathcal{A}| |$ \\
        Edge substitution $e_1\to e_2$ & 0 & $| \epsilon(e_2)/|\mathcal{A}|-\epsilon(e_1)/|\mathcal{A}| | $ \\
        \bottomrule
    \end{tabularx}
    \caption{GED transformation costs}
    \label{tab:GED_costs}
\end{table}

The computation of GED can be customized by acting on the costs of graph transformation operations.
The cost used for the distance analysis follows these rules (see Table~\ref{tab:GED_costs}).
\begin{itemize}
    \item The cost of adding or removing a node is computed as the ratio between the cardinality of the interested node (the number of times an attack step has been used in the report) and the total number of nodes in the graph (the number of attack steps in all the reports).
    \item The cost of adding or removing an edge is computed as the ratio between the cardinality of the specific edge and the total number of edges in the graph.
    \item In case of a substitution, the cost is computed as the absolute value of the difference between the cost of the removed entity and the added one. 
\end{itemize}

\begin{figure}[tb]
    \centering
    \includegraphics[width=0.9\textwidth]{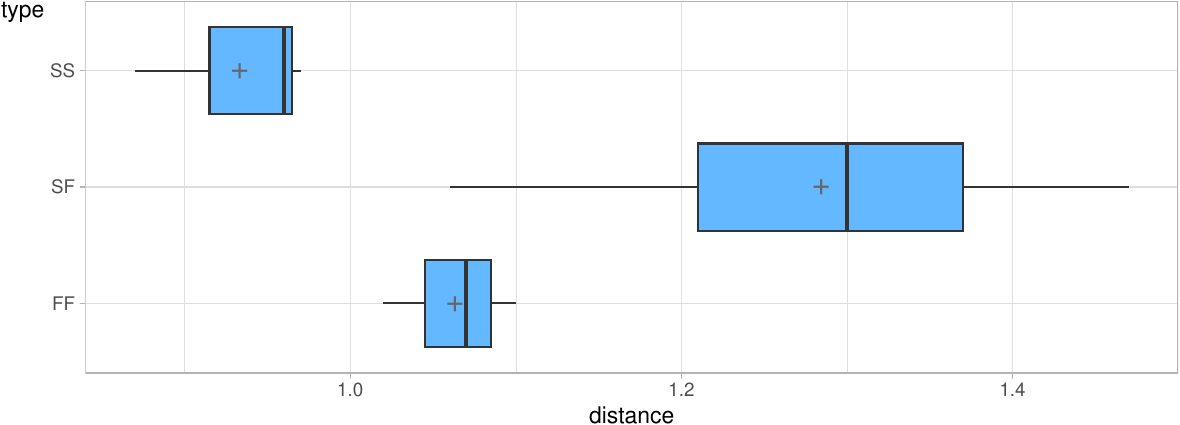}
    \caption{Distribution of distances presented in Tab.\ref{tab:similarity_index}}
    \label{fig:distribution_dist}
\end{figure}

The similarity index computed on process graphs is shown in Tab.~\ref{tab:similarity_index}, divided according to their success. For instance, the process graph obtained from the reports of subjects who successfully tampered with the arithmetic application has a similarity of 0.87 compared to the process graph generated from the reports of subjects who successfully tampered with the number application.
On average, two graphs of successful attack tasks are more similar than two process graphs of failed tasks (see Figure~\ref{fig:distribution_dist}. 
On average, subjects who succeeded followed similar processes (average similarity 0.93), regardless of the treatment, while the processes of the subjects who failed were less similar\footnote{Paraphrasing Tolstoy, ``succeeding subjects are all alike; every failing subject failed in its own way.''} (average similarity 1.07).
Interestingly, the process of successful tasks was clearly different from that of subjects who failed (average similarity of 1.29), highlighting the importance of following the correct process to approach assets. 

\begin{table}[tb]
\caption{Similarity index computed on the graphs depending on the app (tictactoe=ttt, numbers=num, and arithmetic = arith) and the succeeded vs.\ failed tasks (S vs.\ F).}
\centering
\begin{tabular}{lllllll}
\toprule
          & ttt (S) & num (S) & arith (S) & ttt (F) &     num (F) & arith (F) \\
\midrule
ttt (S)   &  -      & 0.96    & 0.97      & 1.29    & 1.37    & 1.43      \\
num (S)   &         & -       & 0.87      & 1.06    & 1.12    & 1.35      \\
arith (S) &         &         & -         & 1.21    & 1.31    & 1.47      \\
ttt (F)   &         &         &           & -       & 1.02    & 1.07      \\
num (F)   &         &         &           &         & -       & 1.1       \\
arith (F) &         &         &           &         &         & -        \\
\bottomrule
\end{tabular}

\label{tab:similarity_index}
\end{table}

The analysis of the processes also revealed that the treatments did not significantly impact the attack strategy.
Since the analysis in Section~\ref{sec:results} showed that tasks were less successful and slower, and the process remained unchanged, we can infer that the application of obfuscation techniques slowed down the processes, and those that did not succeed may have succeeded with more time.
More precisely, understanding the code to locate the proper area to modify becomes more complex.
Hence, this is another potential piece of evidence that the correlation with objective metrics may be the leading factor in delaying attackers. However, further experiments focusing on process analysis would be required to make the findings scientifically relevant.
Perhaps some students were employing a different strategy, which is why they failed on the protected application.

Moreover, we have found some evidence in the reports that a learning process happened. Indeed, some reports used the step name ``{recognize similarity with already analysed protected application}'' and some free text descriptions indicated that they recognized similar patterns from the attack task in a previous period ``the code looked similar to the first program''. 

\changed{Summarizing, we performed an analysis of the attack steps followed by the subjects, which resulted in two additional findings: succeeding subjects typically performed similar attacks, and the used protection did not alter the subjects' attack strategy.}

\end{document}